\documentclass[11pt]{article}

\usepackage{amsmath,amssymb,amsfonts,bbm,bm}
\usepackage{slashed}

\usepackage{tocloft} 

\usepackage[nosort]{cite} 
\usepackage{graphicx,color}
\usepackage[colorlinks=true, linkcolor=blue, citecolor=blue, linktoc=all]{hyperref}

\usepackage[usenames,dvipsnames]{xcolor}

\usepackage{tikz-cd}
\usepackage{pict2e}
\usepackage{physics}
\usepackage{comment} 
\usepackage[makeroom]{cancel}

\newcommand{\beq}{\begin{eqnarray}}
\newcommand{\eeq}{\end{eqnarray}}
\newcommand{\beqn}{\begin{eqnarray}}
\newcommand{\eeqn}{\end{eqnarray}}
\newcommand{\bea}{\begin{eqnarray}}
\newcommand{\eea}{\end{eqnarray}}
\newcommand{\be}{\begin{equation}}
\newcommand{\ee}{\end{equation}}

\usepackage[normalem]{ulem}

\newcommand{\caltech}[1]{
  \centerline{
    \begin{minipage}[c]{0.7\textwidth}
      \begin{center}
      ${}^{#1}$ Walter Burke Institute for Theoretical Physics,\\
      California Institute of Technology, Pasadena, CA 91125, U.S.A.
      \end{center}
    \end{minipage}
  }
}

\usepackage{mathrsfs,mathtools}
\renewcommand\mathbb[1]{\mathbbm{#1}}

\makeatletter
\DeclareRobustCommand{\loplus}{\mathbin{\mathpalette\dog@lsemi{+}}}
\DeclareRobustCommand{\lotimes}{\mathbin{\mathpalette\dog@lsemi{\times}}}
\DeclareRobustCommand{\roplus}{\mathbin{\mathpalette\dog@rsemi{+}}}
\DeclareRobustCommand{\rotimes}{\mathbin{\mathpalette\dog@rsemi{\times}}}

\newcommand{\dog@rsemi}[2]{\dog@semi{#1}{#2}{-90,90}}
\newcommand{\dog@lsemi}[2]{\dog@semi{#1}{#2}{270,90}}
\newcommand{\dog@semi}[3]{%
  \begingroup
  \sbox\z@{$\m@th#1#2$}%
  \setlength{\unitlength}{\dimexpr\ht\z@+\dp\z@\relax}%
  \makebox[\wd\z@]{\raisebox{-\dp\z@}{%
    \begin{picture}(1,1)
    \linethickness{\variable@rule{#1}}
    \roundcap
    \put(0.5,0.5){\makebox(0,0){\raisebox{\dp\z@}{$\m@th#1#2$}}}
    \put(0.5,0.5){\arc[#3]{0.5}}
    \end{picture}%
  }}%
  \endgroup
}
\newcommand{\variable@rule}[1]{%
  \fontdimen8  
  \ifx#1\displaystyle\textfont3\else
    \ifx#1\textstyle\textfont3\else
      \ifx#1\scriptstyle\scriptfont3\else
        \scriptscriptfont3\relax
  \fi\fi\fi
}
\DeclareRobustCommand{\loplus}{\mathbin{\mathpalette\dog@lsemi{+}}}

\usepackage{amsthm}

\usepackage{tikz}
\usepackage{booktabs}
\usepackage{amssymb}
\usetikzlibrary{arrows.meta, positioning, calc}
\renewcommand{\op}[1]{\boldsymbol{#1}}
\usepackage{tabularx}
\usepackage{array}
\usepackage{tcolorbox}
\usepackage{float}

\newcommand{\thistitle}{How to have your wormholes and factorize, too}

\begin{document}

\title{\thistitle}
\author{
	Marc S. Klinger 
	\\
	\\
	{\small \emph{\caltech{}}}
	\\
	}
\date{}
\maketitle
\vspace{-0.5cm}
\begin{abstract}
\vspace{0.3cm}
We review three well known inconsistencies in the standard mathematical formulation of semiclassical gravity: the factorization problem, the information problem, and the closed universe problem. Building upon recent work \cite{Liu:2025ikq}, we explore how modifying the semiclassical holographic dictionary may provide the necessary freedom to resolve these three problems in a unified manner while maintaining more well established aspects of the standard correspondence. Using the modified semiclassical holographic dictionary as a scaffolding, we propose a program for constructing an `extended' semiclassical gravitational path integral which (i) is manifestly factorizing, (ii) computes a von Neumann entropy which satisfies the Page curve, and (iii) incorporates new operators that create closed baby universe states. Our construction may be interpreted as imposing a semiclassical version of background independence/a no global symmetry condition, defining a modified large N limit, preparing an ensemble of dual theories, or enforcing observer rules using gravitational degrees of freedom. 
\end{abstract}

\begingroup
\hypersetup{linkcolor=black}
\tableofcontents
\endgroup
\noindent\rule{\textwidth}{0.6pt}

\setcounter{footnote}{0}
\renewcommand{\thefootnote}{\arabic{footnote}}

\newcommand{\curr}[1]{\mathbb{J}_{#1}}
\newcommand{\constr}[1]{\mathbb{M}_{#1}}
\newcommand{\chgdens}[1]{\mathbb{Q}_{#1}}
\newcommand{\spac}[1]{S_{#1}}
\newcommand{\hyper}[1]{\Sigma_{#1}}
\newcommand{\chg}[1]{\mathbb{H}_{#1}}
\newcommand{\ThomSig}[1]{\hat{\Sigma}_{#1}}
\newcommand{\ThomS}[1]{\hat{S}_{#1}}
\newcommand{\discuss}[1]{{\color{red} #1}}
\theoremstyle{definition}
\newtheorem{theorem}{Theorem}[section]
\newtheorem{example}{Example}[section]
\newtheorem{definition}{Definition}[section]

\section{Introduction}

The last several years have revealed a lot about the fundamental structure of quantum gravity, especially in the semiclassical regime. Although it would be a vast over-generalization, one might argue that these insights have been uncovered by employing three main modes of inquiry -- holography, the gravitational path integral, and quantum information theory. Of course, these three modalities are not entirely or even largely independent. Yet, each offers different advantages that often complement the disadvantages of the other two. To this end, one might hope that by patching together the insights of these different approaches a unified picture of quantum gravity should emerge. However, the interfaces between these ideas are incompatible, revealing many puzzles about the overarching nature of the theory they purport to collectively describe.

The holographic principle dictates that the dynamical degrees of freedom contained in a theory of quantum gravity can be effectively encoded in the degrees of freedom of some lower dimensional quantum mechanical system \cite{Susskind:1994vu}.\footnote{Here and in the remainder of the note we use the term `quantum mechanical' to refer to any system that is quantum in nature but does not include dynamical gravity. In particular, a standard quantum field theory on a curved background is an example of a quantum mechanical system.} The most complete manifestation of the holographic principle is the AdS/CFT correspondence, which claims that the theory of quantum gravity in an asympototically anti de Sitter (AdS) spacetime is in fact equivalent to a non-gravitational theory on the (conformal) boundary of the spacetime itself \cite{Maldacena:1997re}. More generally, we may regard the holographic principle as furnishing a dictionary, $\beta$, which maps between the collection of (diffeomorphism invariant) observables in the quantum gravitational theory, $\mathcal{B}$, and the collection of observables of some holographically dual quantum mechanical system, $\mathcal{A}$. With this dictionary in hand, one can, in principle, compute any quantity of interest in the quantum gravitational theory using tools from standard quantum theory. In the present note, we will not primarily be concerned with any complete, non-perturbative holographic dictionary, but rather a holographic dictionary which adequately describes semiclassical gravity. As such, it will also be useful to discuss what might be referred to as `approximate' holography. Approximate holography refers to a scenario in which an \emph{effective} theory of gravity is encoded in a sequence of standard quantum mechanical theories. A standard example of approximate holography is the large-$N$ limit of the AdS/CFT correspondence. Let $\beta_N: \mathcal{A}_N \rightarrow \mathcal{B}_{G_{\textrm{N}}}$ denote a family of holographic dictionaries mapping between a quantum gravity theory at finite $G_{\textrm{N}}$ -- $\mathcal{B}_{G_{\textrm{N}}}$ -- and an $SU(N)$ super Yang-Mills theory -- $\mathcal{A}_N$. To access semiclassical gravitational features from this correspondence one is instructed to take the limit $G_{\textrm{N}} \rightarrow 0$ which coincides with the limit $N \rightarrow \infty$ on the CFT side.  

In contrast to the holographic approach, which seeks to understand the gravitational theory by translating it into a quantum mechanical theory, the gravitational path integral\footnote{Again, when we refer to the gravitational path integral here, we primarily mean in an appropriate semiclassical limit.} seeks to treat gravity directly on its own terms within the bulk spacetime \cite{Hawking:1978jz}. To this end, while the gravitational path integral bears a strong resemblance to the path integral machinery standard to quantum theory it is necessarily modified in many interesting and sometimes counter-intuitive ways. Let us consider, first, a quantum field theory on a fixed background spacetime $M$. Assuming that $M$ is globally hyperbolic, it can be foliated as $M \simeq I \times \Sigma$ for $I$ a timelike interval and such that $\Sigma_t$ are spacelike Cauchy slices for each $t \in I$. The standard path integral may be regarded, in the language of Atiyah and Segal's axiomatic treatment \cite{Atiyah1988TQFT}, as a map, $Z$, which makes the following assignments.
\begin{enumerate}
	\item To each Cauchy slice, $Z(\Sigma)$ assigns a Hilbert space spanned by the possible initial data which could be specified up to an appropriate polarization.
	\item To each submanifold of spacetime $T \subset M$ with boundaries $\partial T = \Sigma_1 \sqcup \Sigma_2$, $Z(T)$ assigns a space of isometric maps $U: Z(\Sigma_2) \rightarrow Z(\Sigma_1)$.
	\item To each closed submanifold $C \subseteq M$, $Z(C)$ assigns a number which can be interpreted as the partition function of the quantum field theory specified on this manifold.
\end{enumerate}    
These assignments are represented as equations and diagrams in Figure \ref{fig:AtiyahSegal}. More generally, they can all be absorbed into a single rule which is that the path integral takes as input a set of boundary conditions $\textrm{BC}$ (and possibly operator insertions) and outputs a complex number
\beq \label{eq:ZGen}
	Z(\textrm{BC}) = \int \mathscr{D} \Phi \; \mathcal{O}_{\textrm{BC}}[\Phi] e^{-S[\Phi]}. 
\eeq
Depending upon the nature of these boundary conditions e.g. if they are specified on a single surface, multiple surfaces, or left empty, we interpret the resulting complex number as a component of a wavefunction, matrix element of an operator, or a partition function. Naturally, left unmodified this construction would run into several stopping points when applied to quantum gravity \cite{Colafranceschi:2023moh,Iliesiu:2022kny,Balasubramanian:2025jeu,Balasubramanian:2025hns}. The standard trick for circumventing these challenges is the proposal that the gravitational path integral should be used more or less in the same manner as the standard path integral but with the additional dictum that one `sums' over all legal bulk spacetimes which are compatible with the given boundary conditions e.g.
\beq \label{eq:ZGenQG}
	\mathcal{Z}(\textrm{BC}) = \sum_{M \in \mathscr{S}_{\textrm{BC}}} Z_M(\textrm{BC}). 
\eeq
Here, $\mathscr{S}_{\textrm{BC}}$ is the `set' of the geometries which are `compatible' with the given boundary conditions, $Z_M$ is a standard path integral over a fixed bulk $M$, and we have introduced the notation $\mathcal{Z}$ to refer to a gravitational path integral in contrast to $Z$ for a standard path integral. Both of the quoted words in the previous sentence are only vaguely defined. What it truly means to sum over geometries or for such geometries to be compatible with boundary conditions is rather ambiguous.  

\begin{figure}[h]
    \centering
    \begin{minipage}{0.8\textwidth}
        \beq \label{eq:ZHS}
        		\ket{\Psi} \in Z(\Sigma), \qquad \bra{\phi} \ket{\Psi} = \int^{\phi} \mathscr{D} \Phi \; \mathcal{O}_{\Psi}[\Phi] e^{-S[\Phi]} \;=\;
\begin{tikzpicture}[baseline={(0,1.6)}, scale=0.75]

\draw[thick] (-1.2,0) -- (-1.2,3);
\draw[thick] ( 1.2,0) -- ( 1.2,3);

\draw[thick] (-1.2,0) arc (180:360:1.2 and 0.6);

\draw[thick] (-1.2,3) arc (180:360:1.2 and 0.6);
\draw[thick,dashed] (-1.2,3) arc (180:0:1.2 and 0.6);

\node at (0,1.5) {$\mathcal{O}_{\Psi}$};

\node at (0,3.05) {$\phi$};
\node[above] at (0,3.6) {$\Sigma$};

\end{tikzpicture}
        \eeq
        \beq \label{eq:ZIso}
        		U \in Z(T), \qquad \bra{\phi_1} U \ket{\phi_2} = \int_{\phi_2}^{\phi_1} \mathscr{D} \Phi \; \mathcal{O}_U[\Phi] e^{-S[\Phi]} \;=\;
\begin{tikzpicture}[baseline={(0,1.6)}, scale=0.75]

\draw[thick] (-1.2,0) -- (-1.2,3);
\draw[thick] ( 1.2,0) -- ( 1.2,3);

\draw[thick] (-1.2,0) arc (180:360:1.2 and 0.6);
\draw[thick,dashed] (-1.2,0) arc (180:0:1.2 and 0.6);

\draw[thick] (-1.2,3) arc (180:360:1.2 and 0.6);
\draw[thick,dashed] (-1.2,3) arc (180:0:1.2 and 0.6);

\node at (0,1.5) {$\mathcal{O}_{U}$};

\node at (0,3.05) {$\phi_1$};
\node[above] at (0,3.6) {$\Sigma_1$};

\node at (0,-0.05) {$\phi_2$};
\node[below] at (0,-0.6) {$\Sigma_2$};

\end{tikzpicture}
        \eeq
        \beq \label{eq:ZPart}
        		Z(M) = \int \mathscr{D}\Phi \; e^{-S[\Phi]} \;=\;
\begin{tikzpicture}[baseline={(0,0)}, scale=0.75]

\draw[thick] (0,0) circle (1.4);

\draw[thick] (-1.4,0) arc (180:360:1.4 and 0.5);
\draw[thick,dashed] (-1.4,0) arc (180:0:1.4 and 0.5);

\node at (0,0) {$M$};

\end{tikzpicture}
        \eeq
    \end{minipage}
    \caption{The standard path integral of quantum field theory. The notation $\int_{\phi_2}^{\phi_1} \equiv \int_{\{\Phi \in \Gamma(M,E) \; | \; \Phi\rvert_{\Sigma_1} = \phi_1, \Phi\rvert_{\Sigma_2} = \phi_2\}}$ indicates the set of all field configurations satisfying boundary conditions $\phi_1$ at the upper boundary and $\phi_2$ at the lower boundary. If either of the boundary conditions are left empty, one should integrate over the space of fields unconstrained on the given boundary besides necessary fall-off conditions.}
    \label{fig:AtiyahSegal}
\end{figure}

Finally, we have the quantum information theoretic approach. For the purposes of the discussion in this paper, we will primarily regard the quantum information theoretic approach as synonymous with the notion that  spacetime can be understood as an `emergent' phenomenon quantified in terms of a (possibly only approximate) quantum error correcting code \cite{VanRaamsdonk:2010pw,Almheiri:2014lwa,Almheiri:2020cfm,Cotler:2017erl}. The general set up of this picture is as follows: We have two Hilbert spaces, $\mathscr{H}_{\textrm{fund.}}$ and $\mathscr{H}_{\textrm{eff.}}$, which describe, respectively, the `fundamental' and `effective' degrees of freedom of a particular system. The effective degrees of freedom are `encoded' into the fundamental Hilbert space via a linear map $V: \mathscr{H}_{\textrm{eff.}} \rightarrow \mathscr{H}_{\textrm{fund.}}$. In some sense, the goal of the emergent spacetime program is to `invert' the encoding map so as to deduce how features like smooth geometry -- which one would not expect to be associated with typical quantum gravitational microstates -- emerge from the judicious superposition of those microstates. Of course, the emergent spacetime picture is quite compatible with the holographic principle with two main caveats. First, the holographic principle is intended to be a fully non-perturbative, exact duality between two theories. Conversely, the effective states in the quantum error correcting picture are not expected and indeed should not be in one to one correspondence with the fundamental microstates. In this sense, the emergent spacetime story is more in line with the `approximate' version of the holographic correspondence. This is quite appropriate for our investigation of the semiclassical limit. For instance, in the language of AdS/CFT, $\mathscr{H}_{\textrm{fund.}}$ can be identified with the complete Hilbert space of a holographic CFT while $\mathscr{H}_{\textrm{eff.}}$ is the Hilbert space of perturbative, geometric states. The second main difference between the quantum information theoretic point of view and the holographic point of view is the kind of things that one is interested in computing. The quantum information theoretic point of view places a great emphasis on the relationship between geometric quantities in the effective description and entropic/entanglement related quantities to do with the encoding map, the inversion map, or the fundamental Hilbert space.  

As we have already emphasized, the distinction between these three perspectives is rather artificial. It is hopefully clear from the above discussion that one typically transitions naturally between the three. For instance, the holographic correspondence can be understood as an equivalence between the gravitational path integral and a standard quantum mechanical path integral on some fixed, typically lower dimensional manifold. It can also be understood as a limit of the approximate quantum error correcting picture wherein the effective gravitational description becomes a bulk microstate description. Likewise, the quantum information theoretic approach has been aided substantially by the gravitational path integral. It was in this context that the correspondence between extremal surfaces and the generalized gravitational entropy was first understood \cite{Ryu:2006bv}. With all this being said, however, we have emphasized the distinction between these three different points of view because the interfaces between them reveal obstructions that obscure our view of quantum gravity. 

\begin{figure}[ht]
\centering
\begin{tikzpicture}[
    every node/.style={align=center},
    main/.style={font=\bfseries},
    arrow/.style={-{Latex[length=3mm,width=2mm]}, thick}
]

\node[main] (H) at (90:4) {Holographic\\ Principle};
\node[main] (G) at (210:4) {Gravitational\\ Path Integral};
\node[main] (Q) at (330:4) {Quantum\\ Information};

\coordinate (HGmid) at ($(H)!0.5!(G)$);
\coordinate (GQmid) at ($(G)!0.5!(Q)$);
\coordinate (HQmid) at ($(H)!0.5!(Q)$);

\node (F) at (HGmid) {The Factorization\\ Problem};
\node (I) at (GQmid) {The Information\\ Problem};
\node (L) at (HQmid) {The Closed Universe\\ Problem};

\draw[arrow] (H) -- (F);
\draw[arrow] (H) -- (L);

\draw[arrow] (G) -- (F);
\draw[arrow] (G) -- (I);

\draw[arrow] (Q) -- (I);
\draw[arrow] (Q) -- (L);

\end{tikzpicture}
\caption{Obstructions to the consistency of semiclassical gravity.}
\label{fig:QGTriangle}
\end{figure}

In section~\ref{sec: obst} we delve into three well known incompatibilities between the approaches we have introduced. These incompatibilities are summarized in Figure \ref{fig:QGTriangle}. We begin in subsection~\ref{subsec: H-GPI}, in which we identify an incompatibility between the holographic principle and the gravitational path integral. Namely, the holographic principle dictates that the semiclassical gravity theory should be dual (at least in some appropriate limiting sense) to a standard quantum theory. On the other hand, wormhole contributions to the gravitational path integral imply that it cannot reproduce the expected factorization properties of a standard quantum theory. This is the \emph{Factorization Problem} \cite{Witten:1999xp,Maldacena:2004rf}. In subsection~\ref{subsec: GPI-QI}, we identify an incompatibility between the gravitational path integral and insights from the quantum information theoretic approach to quantum gravity. Based on quantum information theoretic arguments, one expects the von Neumann entropy of Hawking radiation to obey a Page curve, as would imply that the quantum gravitational theory is consistent with unitarity. Naively, computing the von Neumann entropy via the gravitational path integral using the replica trick appears to satisfy this expectation \cite{Penington:2019kki,Almheiri:2019qdq}. However, the precise definition of the replica trick reveals an incompatibility when the non-factorization of the gravitational path integral is taken into account. In particular, the von Neumann entropy of a state defined by the gravitational path integral is determined by a gravitational path integral in which the contribution of replica wormholes are subtracted away. This suggests that the entropy of Hawking radiation actually satisfies a Hawking, rather than a Page curve.\footnote{We should emphasize that this is consistent with the resolutions described in \cite{Penington:2019kki,Almheiri:2019qdq}. We are placing an additional constraint on the information problem which is that the quantity appearing in the Page curve should be the von Neumann entropy of a single state rather than the average entropy over an ensemble of states.} This is (our form of) the \emph{Information Problem}. Finally, in subsection~\ref{subsec: QI-H} we identify an incompatibility between the approximate error correcting picture of emergent gravity and the holographic principle. This is best illustrated through the AdS/CFT correspondence and so we restrict our attention to this case. The nature of the large N limit used to map used to analyze semiclassical gravity prohibits the emergence of states which are mixed with respect to the algebra of large N single trace operators in the dual CFT. This implies that the standard holographic dictionary cannot accommodate the emergence of non-trivial closed universes \cite{Gesteau:2025obm,Antonini:2024mci}. Alternatively, assuming that the standard large N dictionary cannot be meaningfully altered, this suggests that closed universes are described by a one-dimensional Hilbert space \cite{McNamara:2020uza}. This is the \emph{Closed Universe Problem}. 

The main argument of the present note is that these three problems share a single common resolution, which we present in section~\ref{sec: res}. Following the lead of recent work \cite{Liu:2025cml,Liu:2025ikq}, we argue that the holographic dictionary must be modified. Again, focusing on the AdS/CFT correspondence, the dependence of the gravitational theory on $G_{\textrm{N}}$ as a parameter is smooth, while the dependence of the dual CFT on $N$ is erratic. Consequently, the limits $G_{\textrm{N}} \rightarrow 0$ and $N \rightarrow \infty$ are not compatible and the equality of these limits cannot be true. To rectify this problem, \cite{Liu:2025ikq} proposed that the semiclassical gravity theory described by the gravitational path integral must be dual to a subtheory of the full holographic CFT which depends smoothly on the parameter $N$. To formulate this modified holographic correspondence, we introduce a projection map $E$ which isolates the smooth part of the CFT. We then build beyond the proposal of \cite{Liu:2025ikq} by arguing that the projection $E$ encodes some non-perturbative data about the dual theory which can be used to modify the gravitational theory. We propose a set of conditions under which this modified gravitational theory can be constructed such that it (i) factorizes (section~\ref{subsec: WHRenorm}), (ii) gives rise to a von Neumann entropy for states which satisfies the Page curve (section~\ref{subsec: condEnt}), and (iii) allows for the emergence of non-trivial closed universe physics (section~\ref{subsec: SSandBU}). 

We conclude in section~\ref{sec: discuss} in which we briefly compare our approach to other related ideas in the literature. To improve the readability of the note, we have relegated most involved mathematical discussions into the appendix. The main text can largely be read without consulting these appendices, but they provide very useful background for understanding some of the more technical aspects of our analysis. In Appendix \ref{app: QCon}, we give a comprehensive overview of quantum conditional probability theory. This reviews a series of recent work \cite{AliAhmad:2024saq,AliAhmad:2025oli,AliAhmad:2025bnd,Klinger:2025tvg} which has developed the role of non-commutative conditional expectations and related maps in organizing the structure and information content of quantum theories. Especially, we emphasize the structure theory of algebraic inclusions admitting operator valued weights, and the factorization of the von Neumann entropy for states constructed using generalized conditional expectations. In Appendix \ref{app: AlgPI}, we describe how one can construct a path integral given a rather generic operator algebra provided it admits a special form of representation. This provides an important tool since it allows for us to move back and forth between path integral and operator algebraic manipulations.  

\section{Obstructions to the consistency of semiclassical gravity} \label{sec: obst}

\subsection{Between Holography and the Path Integral: The Factorization Problem} \label{subsec: H-GPI}

The first obstruction is the factorization problem, which underscores an incompatibility between the holographic principle and the gravitational path integral. According to the holographic principle, a theory of quantum gravity should be dual to a standard quantum theory. Thus, given a quantum gravity theory formulated in terms of a gravitational path integral $\mathcal{Z}$, there should exist a quantum theory formulated in terms of a standard path integral $Z$ along with a holographic dictionary, $\beta$, assigning to any boundary conditions $\textrm{BC}$ in the gravitational theory a dual space $M^{\beta}$ and a set of boundary conditions $\textrm{BC}^{\beta}$ such that
\beq \label{Z Holographic Dictionary}
	\mathcal{Z}(\textrm{BC}) = Z(\textrm{BC}^{\beta}). 
\eeq

The obstruction emerges when one considers eqn. \eqref{Z Holographic Dictionary} applied to the disjoint union of manifolds. At the co-dimension zero level, e.g. for partition functions, one expects a standard path integral to factorize as
\beq \label{QMPI1}
	Z(\bigsqcup_{i = 1}^n M_i) = \prod_{i = 1}^n Z(M_i). 
\eeq
Likewise, at co-dimension one, one expects the Hilbert space assignment to factorize as
\beq \label{QMPI2}
	Z(\bigsqcup_{i = 1}^n \Sigma_i) = \bigotimes_{i = 1}^n Z(\Sigma_i). 
\eeq
However, if we consider the analogous assignment in the gravitational path integral neither of these factorizations seem assured. For partition functions, $\mathcal{Z}(\bigsqcup_{i = 1}^n M_i)$ involves a sum over all bulk spacetimes compatible with the associated `closed boundary' conditions. This is likewise true for $\mathcal{Z}(\bigsqcup_{i = 1}^n \Sigma_i)$ which can include contributions from bulk spacetimes that connect the slices $\Sigma_i$. Thus, at least schematically, we find
\beq \label{QGPIFact}
	\mathcal{Z}(\bigsqcup_{i = 1}^n M_i) = \bigg(\prod_{i = 1}^n \mathcal{Z}(M_i)\bigg) \mathcal{Z}_{\textrm{conn.}}(\bigsqcup_{i = 1}^n M_i), \qquad \mathcal{Z}(\bigsqcup_{i = 1}^n \Sigma_i) = \bigg(\bigotimes_{i = 1}^n \mathcal{Z}(\Sigma_i)\bigg) \otimes \mathcal{Z}_{\textrm{conn.}}(\bigsqcup_{i = 1}^n \Sigma_i). 
\eeq
Here, $\mathcal{Z}_{\textrm{conn.}}$ is the `connected' contribution to the gravitational path integral. Assuming that the holographic dictionary distributes as
\beq \label{factorization assumption}
	(\bigsqcup_{i = 1}^n M_i)^{\beta} = \bigsqcup_{i = 1}^n M^{\beta}_i, \qquad (\bigsqcup_{i = 1}^{n} \Sigma_i)^{\beta} = \bigsqcup_{i = 1}^{n} \Sigma^{\beta}_i,
\eeq 
there is an obvious (possibility for) contradiction between \eqref{QMPI1}, \eqref{QMPI2} and \eqref{QGPIFact}:
\begin{flalign} \label{factorization codim 0}
	\prod_{i = 1}^n Z(M^{\beta}_i) &= Z(\bigsqcup_{i = 1}^n M_i^{\beta}) \nonumber \\
	 &= \mathcal{Z}(\bigsqcup_{i = 1}^n M_i) \nonumber \\
	 &= \bigg(\prod_{i = 1}^n \mathcal{Z}(M_i)\bigg) \mathcal{Z}_{\textrm{conn.}}(\bigsqcup_{i = 1}^n M_i) = \bigg(\prod_{i = 1}^n Z(M^{\beta}_i)\bigg) \mathcal{Z}_{\textrm{conn.}}(\bigsqcup_{i = 1}^n M_i). 
\end{flalign}
An analogous sequence of `equalities' hold at co-dimension one:
\begin{flalign} \label{factorization codim 1}
	\bigotimes_{i = 1}^n Z(\Sigma_i^{\beta}) &= Z(\bigsqcup_{i = 1}^n \Sigma_i^{\beta}) \nonumber \\
	&= \mathcal{Z}(\bigsqcup_{i = 1}^n \Sigma_i^{\beta}) \nonumber \\
	&= \bigg(\bigotimes_{i = 1}^n \mathcal{Z}(\Sigma_i^{\beta})\bigg) \otimes \mathcal{Z}_{\textrm{conn.}}(\bigsqcup_{i = 1}^n \Sigma_i) = \bigg(\bigotimes_{i = 1}^n Z(\Sigma_i^{\beta})\bigg) \otimes \mathcal{Z}_{\textrm{conn.}}(\bigsqcup_{i = 1}^n \Sigma_i). 
\end{flalign}
Clearly, eqn. \eqref{factorization codim 0} and \eqref{factorization codim 1} can only be true equations if $\mathcal{Z}_{\textrm{conn.}}$ is trivial.\footnote{In principle, one could also point to \eqref{factorization assumption} as the culprit rather than the non-trivial nature of $\mathcal{Z}_{\textrm{conn.}}$. As we will see, however, the correct resolution to the factorization problem will require a modification to the holographic dictionary which could have been absorbed either into a modification of \eqref{Z Holographic Dictionary} leaving \eqref{factorization assumption} unchanged, or visa-versa. For our purposes, it will be more natural to leave \eqref{factorization assumption} fixed and modify \eqref{Z Holographic Dictionary}.} 
In a nutshell, this is the factorization problem -- at the semiclassical level the `connected' contribution to the partition function is not zero.  

\subsection{Between the Path Integral and Quantum Information: The Information Problem} \label{subsec: GPI-QI}

The second, and quite closely related, obstruction is the information problem, which underscores an incompatibility between the gravitational path integral and notions from quantum information. To understand the information problem, let us briefly review how the path integral can be used to compute quantum information theoretic quantities, like the entanglement entropy. For the purpose of this discussion we will dispense with more technical concerns such as the well definedness of the von Neumann entropy for states on subregion algebras in local quantum field theories.\footnote{For instance, the reader may interpret all of the following quantites as being suitably regularized to avoid standard divergences.} 

Consider a standard quantum field theory defined in a bulk spacetime $M$. Let $\Sigma$ be a complete Cauchy slice in $M$. Using the equations depicted in Figure \ref{fig:AtiyahSegal}, we can write down the expectation value of any operator $U \in Z(T)$ where $\partial T = \Sigma \cup \Sigma$ as
\beq \label{PI Exp}
	\bra{\Psi} U \ket{\Psi} = \int \mathscr{D} \Phi \; \overline{\mathcal{O}_{\Psi}[\Phi]} \mathcal{O}_U[\Phi] \mathcal{O}_{\Psi}[\Phi] e^{-S[\Phi]} = \int \mathscr{D} \Phi \; |\mathcal{O}_{\Psi}[\Phi]|^2 \mathcal{O}_{U}[\Phi] e^{-S[\Phi]}. 
\eeq
This should be visualized as preparing the state $\Psi$ and its conjugate and then propagating them through the bulk with the operator $U$.\footnote{The last equality is justified since the path integral includes an implicit time ordering. So insertions can be manipulated as Abelian functions in the path integral as long as the proper operator ordering is taken at the end, remembering to treat $|\mathcal{O}_{\Psi}[\Phi]|^2 = \overline{\mathcal{O}_{\Psi}[\Phi]} \mathcal{O}_{\Psi}[\Phi]$.} The full collection of bounded operators on the Hilbert space $Z(\Sigma)$, forms the algebra $\mathscr{B}(Z(\Sigma))$. To each subregion $R \subset \Sigma$ we can associate a subalgebra $\mathscr{A}_R \subseteq \mathscr{B}(Z(\Sigma))$, which consists of those operators with support within the given subregion. Now, let us partition the field $\Phi = (\Phi_R, \Phi_{R^c})$, where $\Phi_{R}$ is supported in the region $R$ and $\Phi_{R^c}$ is supported in the complementary region. If $a \in \mathscr{A}_{R}$, its associated insertion will be independent of $\Phi_{R^c}$ and thus the expectation value \eqref{PI Exp} yields
\beq
	\bra{\Psi} a \ket{\Psi} = \int \mathscr{D} \Phi_R \bigg(\int \mathscr{D} \Phi_{R^c} |\mathcal{O}_{\Psi}[\Phi_R,\Phi_{R^c}] |^2 e^{-S[\Phi_R,\Phi_{R^c}]}\bigg) \mathcal{O}_a[\Phi_R] \equiv \int \mathscr{D} \Phi_R \mathcal{O}_a[\Phi_R] e^{-S_{\psi_R}[\Phi_R]}. 
\eeq
Here, we have defined by
\beq
	e^{-S_{\psi_R}[\Phi_R]} \equiv \int \mathscr{D} \Phi_{R^c} |\mathcal{O}_{\Psi}[\Phi_R,\Phi_{R^c}]|^2 e^{-S[\Phi_R,\Phi_{R^c}]}
\eeq
the effective action associated with the `partial trace' of $\ket{\Psi}$ in the complementary region $R^c$. More rigorously, 
\beq \label{Algebraic State from PI}
	\psi_R(a) \equiv \int \mathscr{D} \Phi_R \; \mathcal{O}_a[\Phi_R] e^{-S_{\psi_R}[\Phi_R]}
\eeq
defines a state on the algebra $\mathscr{A}_{R}$. 

On account of the partial trace, the state $\psi_R$ is generically mixed. Heuristically, we can regard
\beq
	\psi_R(a) = \tau_R(\rho_{\psi_R} a) = Z(\textrm{BC}_{\Psi,a}),
\eeq	
where $\tau_R$ is a trace on the algebra $\mathscr{A}_R$. In the last equality, we've emphasized that the number $\psi_R(a)$ can be obtained from a standard path integral computation with an insertion $\textrm{BC}_{\Psi,a}$ which depends on the chosen state and the operator $a$. Since $\psi_R$ is a mixed state, it will have a non-trivial von Neumann entropy:
\beq \label{vNEnt}
	S(\psi_R) \equiv -\tau_R(\rho_{\psi_R} \log \rho_{\psi_R}). 
\eeq	
As written in \eqref{vNEnt}, the entropy is not easily computed from the path integral. To overcome this difficulty, we can employ the replica trick:
\beq
	S(\psi_R) = \lim_{n \rightarrow 1} \frac{1}{1-n} \log \tau_R(\rho_{\psi_R}^n) = \lim_{n \rightarrow 1} \frac{1}{1-n} \log \psi_R^{\otimes n}(s_n^R).
\eeq
Here, $\psi_R^{\otimes n}$ is to be interpreted as a state on the $n$-fold tensor product algebra $\mathscr{A}_R^{\otimes n}$ and $s_n^R$ is the $n$-fold swap operator associated with this algebra. Using \eqref{QMPI1} and \eqref{QMPI2}, we can write
\beq \label{PIvNEnt}
	\psi_R^{\otimes n}(s_n^R) = Z(\bigsqcup_{i = 1}^n M, \textrm{BC}_{\Psi^{\otimes n}, s_n^R}). 
\eeq
The quantity on the right-hand side is a standard path integral over an $n$-replicated spacetime with appropriate insertions for preparing $\Psi$ on each copy and replica boundary conditions for implementing the swap operators. Thus, we obtain a path integral expression for the von Neumann entropy as
\beq
	S(\psi_R) = \lim_{n \rightarrow 1} \frac{1}{1-n} Z(\bigsqcup_{i = 1}^n M, \textrm{BC}_{\Psi^{\otimes n}, s_n^R}). 
\eeq	

One would similarly like to use the gravitational path integral to compute the entropy of states in subregions, at least semiclassically. Naively, we might expect that \eqref{PIvNEnt} simply carries over to the gravitational context so that
\beq \label{NaiveGPIEnt}
	\mathcal{S}(\psi_R) \stackrel{?}{=} \lim_{n \rightarrow 1} \frac{1}{1-n} \mathcal{Z}(\bigsqcup_{i = 1}^n M, \textrm{BC}_{\Psi^{\otimes n}, s_n^R}),
\eeq	
where we have used the notation $\mathcal{S}$ to refer to the von Neumann entropy on the appropriate gravitational algebra associated with $\mathcal{Z}$. Indeed, \eqref{NaiveGPIEnt} has the added allure that it reproduces the expected Page curve when $R$ coincides with the region outside of an evaporating black hole \cite{Penington:2019kki,Almheiri:2019qdq}. However, once again \eqref{QGPIFact} rains on our parade, only this time for the opposite reason. The non-factorization of the gravitational path integral implies that
\beq
	\mathcal{Z}(\bigsqcup_{i = 1}^n M, \textrm{BC}_{\Psi^{\otimes n}, s_n^R}) = \psi_R^{\otimes n}(s_n^R) \mathcal{Z}_{\textrm{conn.}}(\bigsqcup_{i = 1}^n M, \textrm{BC}_{\Psi^{\otimes n}, s_n^R}).
\eeq
Thus, simply by definition, \eqref{NaiveGPIEnt} does not compute the von Neumann entropy of the state $\psi_R$ defined by the gravitational path integral. The correct expression
\beq
	\mathcal{S}(\psi_R) = \lim_{n \rightarrow 1} \frac{1}{1-n} \bigg[ \log\bigg(\mathcal{Z}(\bigsqcup_{i = 1}^n M, \textrm{BC}_{\Psi^{\otimes n},s_n^R})\bigg) - \log\bigg(\mathcal{Z}_{\textrm{conn.}}(\bigsqcup_{i = 1}^n M, \textrm{BC}_{\Psi^{\otimes n},s_n^R})\bigg) \bigg]
\eeq
explicitly subtracts off the connected contribution to the gravitational path integral and therefore reproduces the dreaded Hawking curve. 

One resolution to this tension is to argue that, although it is not equal to the von Neumann entropy of any single state, the quantity \eqref{NaiveGPIEnt} computes the average entropy over an ensemble of theories. What's more, such a quantity may be argued to be the preferred quantum information theoretic measure for understanding the evaporation of a black hole, see e.g. \cite{Renner:2021qbe}. For our purposes, however, we will demand something stronger e.g. a gravitational theory which constructs states with entropy satisfying the Page curve \emph{without} ensemble averaging. From this point of view, we observe an interesting parallel: In the factorization puzzle, the appearance of a non-trivial wormhole contribution violated the expectations of holography and therefore it appeared we should desire that $\mathcal{Z}_{\textrm{conn}.} = 0$. In the information puzzle, the definition of the von Neumann entropy requires we get rid of the connected contribution, but now we want it back!

\subsection{Between Quantum Information and Holography: The Closed Universe Problem} \label{subsec: QI-H}

The final obstruction we will discuss pertains to the physics of closed universes under the standard holographic dictionary. Our discussion in this section is deliberately non-exhaustive; this subject has received a great deal of attention in recent work \cite{Gesteau:2025obm,Antonini:2024mci,Liu:2025cml,Liu:2025ikq,Klinger:2025tvg,Harlow:2025pvj,Antonini:2025ioh,Sasieta:2025vck,Kudler-Flam:2025cki,Abdalla:2025gzn,Engelhardt:2025azi,Akers:2025ahe,Higginbotham:2025clp}. Our main goal is to illustrate the sense in which this obstruction reveals incompatibilities between the emergent spacetime picture and holography. 

The general set up we will employ for analyzing this obstruction is adapted from \cite{Liu:2025cml}. According to the standard (asymptotic) AdS/CFT dictionary, the Hilbert space of a semiclassical gravity theory around a fixed, asymptotically AdS background can be obtained as the limit of a sequence of CFT Hilbert spaces each equipped with a preferred vector state, $\mathscr{H}_{\Psi} \equiv \{\mathscr{H}_N, \ket{\Psi_N}\}_{N \in \mathbb{N}}$. To keep with the path integral oriented presentation we have used up to this point, we can regard each $\mathscr{H}_N = Z_N(\Sigma)$, where $Z_N$ is a CFT path integral and $\ket{\Psi_N}$ as a state prepared by e.g. \eqref{PI Exp}. The collection of physically allowed observables associated with $\mathscr{H}_{\Psi}$ consists of those sequences of operators $A \equiv \{A_N \in B(\mathscr{H}_N)\}_{N \in \mathbb{N}}$ with finite limiting expectation value in the sequence of preferred states $\ket{\Psi_N}$:
\beq \label{eq:stateonlimalg}
	\lim_{N \rightarrow \infty} \bra{\Psi_N} A_N \ket{\Psi_N} < \infty. 
\eeq
We assume that this collection of operators forms an algebra, which we denote by $\mathcal{A}_{\Psi}$. By construction, eqn. \eqref{eq:stateonlimalg} defines a state, $\omega_{\Psi}$, on this algebra, and thus we can perform a GNS construction to obtain a Hilbert space which we also denote by $\mathscr{H}_{\Psi}$. 

The algebra $\mathcal{A}_{\Psi}$ depends explicitly upon the state $\omega_{\Psi}$, and, in turn, upon the full sequence $\{\ket{\Psi_N}\}_{N \in \mathbb{N}}$. However, one might expect the existence of an intrinsically defined algebra $\mathcal{S}$ which possesses a good large $N$ limit in any state. In the AdS/CFT context, this algebra coincides with the large $N$ algebra of single trace operators arising from the sequence of CFTs. Thus, in general, the large $N$ limit constructs an inclusion $\mathcal{S} \subset \mathcal{A}_{\Psi}$. In \cite{Antonini:2023hdh}, a  state $\omega_{\Psi}$ was introduced and proposed to be dual to a geometry in which two (asymptotically) AdS spaces are entangled across a closed universe. This necessarily implies that $\omega_{\Psi}$ defines a \emph{mixed state} on the algebra $\mathcal{S}$ such that the inclusion $\mathcal{S} \subset \mathcal{A}_{\Psi}$ is strict. 

However, it was later argued by Gesteau in \cite{Gesteau:2025obm} that the standard holographic dictionary disallows the emergence of such a mixed state. Consequently, we seem to be faced with two options: the expected closed universe fails to emerge, or the closed universe emerges but its Hilbert space is one-dimensional. The latter perspective is resounded by the observation that the `gravitational entropy' in eqn. \eqref{NaiveGPIEnt} returns an answer of zero when applied to a closed universe.\footnote{As with the Page curve computation, and as we will discuss in detail in the sequel, the failure of the quantity \eqref{NaiveGPIEnt} to be realized as the von Neumann algebra of a single state casts doubt on the validity of the one-dimensional Hilbert space derived in this way.} This is the tension between the emergent spacetime picture and the holographic dictionary: the latter apparently cannot accommodate the emergence of non-trivial closed universes. 

\section{A Resolution: The Modified Semiclassical Holographic Dictionary} \label{sec: res}

The incompatibilities identified in the previous section obstruct the conclusion that our three different approaches are in fact describing the same underlying physical system. In this section, we will argue that it is possible to resolve these inconsistencies by modifying the holographic dictionary. 

To motivate our resolution let us first describe recent work by Liu which has suggested that the gravitational path integral should be regarded as dual to a `projected' version of the holographic CFT \cite{Liu:2025ikq}. According to the standard the AdS/CFT correspondence, we should recover a duality of the form
\beq \label{eq:asymHolo}
	\lim_{G_{\textrm{N}} \rightarrow 0} \mathcal{Z}_{G_{\textrm{N}}} = \lim_{N \rightarrow \infty} Z_N,
\eeq
where we imagine that at each finite $G_{\textrm{N}}$ and $N$ there is an exact duality between the gravity theory, $\mathcal{Z}_{G_{\textrm{N}}}$, and the CFT, $Z_N$. However, the parametric dependence of the gravity theory on $G_N$ is smooth, while the dependence of the CFT on $N$ is erratic. As such, the two limits appearing in eqn. \eqref{eq:asymHolo} are incompatible. To rectify this incompatibility, Liu proposed that the correspondence be modified to the form
\beq \label{eq:modAsymHolo}
	\lim_{G_{\textrm{N}} \rightarrow 0} \mathcal{Z}_{G_{\textrm{N}}} = \lim_{N \rightarrow \infty} Z_N^{\textrm{S}},
\eeq
where $Z_N^{\textrm{S}}$ is a consistent subtheory of the full CFT which depends smoothly on the parameter $N$. 

We can now introduce our proposal. In the following, $\mathcal{Z}$ will denote a (semiclassical) gravitational path integral and $Z$ a putative dual quantum theory. First, we assume that there exists a consistent subtheory $Z^{\textrm{S}} \subset Z$. The inclusion here is meant to indicate that $Z^{\textrm{S}}$ can be used to perform a subset of the computations which are accessible to the complete theory, $Z$. In other words, $Z^{\textrm{S}}$ has as its domain a subclass of boundary conditions relative to $Z$. Of course, this implies that $Z^{\textrm{S}}(\Sigma)$ is a subspace in the conventional sense relative to $Z(\Sigma)$. To simplify our discussion moving forward, let us introduce the notation $\mathfrak{B}_Z$ to denote the abstract set of boundary conditions for the path integral $Z$. In this language, we can simply regard $Z^{\textrm{S}} \subset Z$ as the statement that $\mathfrak{B}_{Z^{\textrm{S}}} \subset \mathfrak{B}_Z$ is a genuine set inclusion and $Z^{\textrm{S}} \equiv Z\rvert_{\mathfrak{B}_{Z^{\textrm{S}}}}$. A good heuristic picture to keep in the back of one's mind is the case in which $Z$ describes a theory of the fields $(\Phi_S, \Phi_E)$ and the subtheory $Z^{\textrm{S}}$ describes only the field $\Phi_S$. Given the inclusion $Z^{\textrm{S}} \subset Z$, we next postulate the existence of a map $E: Z \rightarrow Z^{\textrm{S}}$. Again, this should really be regarded as a mapping at the level of the boundary conditions which are inserted into the path integral. That is $E: \mathfrak{B}_Z \rightarrow \mathfrak{B}_{Z^{\textrm{S}}}$ assigns to each $\textrm{BC} \in \mathfrak{B}_Z$ a restriction $E(\textrm{BC}) \in \mathfrak{B}_{Z^{\textrm{S}}}$. For instance, starting from the boundary condition $\textrm{BC} = (\Phi_S\rvert_{\Sigma} = \phi_S, \Phi_E\rvert_{\Sigma} = \phi_E)$, $Z \circ E(\textrm{BC})$ could retain the boundary condition on $\Phi_S$ but integrate over the unconstrained field $\Phi_E$. We will sometimes use the notation $E(Z)$ to denote the path integral $Z \circ E$ defined on the subset $\mathfrak{B}_{Z^{\textrm{S}}}$. 

With these notations in place, we can introduce the following modified holographic duality:
\beq \label{eq:ModHol}
	\mathcal{Z} = E(Z) \iff \mathcal{Z}(\textrm{BC}) = Z \circ E(\textrm{BC}^{\beta}).
\eeq 
Here, $\beta$ is the holographic dictionary employed in section~\ref{subsec: H-GPI}. At this stage, eqn. \eqref{eq:ModHol} is essentially equivalent to the proposed holographic duality introduced in \cite{Liu:2025ikq}.\footnote{We should emphasize, however, that the notion of filter here is more fine grained than the one introduced in \cite{Liu:2025ikq}. The filter in \cite{Liu:2025ikq} is a map between partition functions which does not presuppose a subdivision at the level of boundary conditions.} There, it is further argued that to ensure eqn. \eqref{eq:ModHol} is consistent with the well verified properties of the standard holographic dictionary the map $E$ must be a positive, linear projection. That is, 
\begin{enumerate}
	\item \textbf{Positive:} $Z(\textrm{BC}) > 0 \implies Z \circ E(\textrm{BC}) > 0$,
	\item \textbf{Linear:} $E(a_1 \textrm{BC}_1 + a_2 \textrm{BC}_2)) = a_1 E(\textrm{BC}_1) + a_2 E(\textrm{BC}_2)$,
	\item \textbf{Projection:} $E \circ E = E$.
\end{enumerate}
The main contribution of the present note is to describe how we can `invert' the projection in an appropriate statistical sense in order to arrive at an `extended' semiclassical gravity theory $\mathcal{Z}^{\textrm{ext.}}$. The form of this theory is determined in part by the `filter' $E$, but also requires additional input. To constrain the form of $E$ and this additional input, we will present a sequence of conditions motivated by the desire to resolve the inconsistencies identified in section~\ref{sec: obst}. This leads to a kind of `bootstrap' in which the extended gravitational theory is determined simply from the small UV input of $E$ and mathematical consistency conditions. 

To construct the extended gravitational path integral, it turns out to be useful to first translate the above construction into the language of operator algebras. In section~\ref{subsec: H-GPI} we have provided a description of how the path integral can be used to compute the matrix elements of a certain class of operators. Conversely, in Appendix~\ref{app: AlgPI} we demonstrate how one can begin with an algebra of operators and \emph{construct} a path integral by choosing a special kind of representation. As such, given the path integral $Z$ we can identify an algebra of operators $\mathscr{A}_Z$. Using this fact, the inclusion $Z^{\textrm{S}} \subset Z$ can be translated into an algebraic inclusion $\mathscr{A}_S \subset \mathscr{A}_Z$. From this point of view, the map $E$ is naturally translated into a map $E: \mathscr{A}_Z \rightarrow \mathscr{A}_S$, satisfying the following properties
\begin{enumerate}
	\item \textbf{Positive:} $E(a^* a) > 0, \qquad \forall a \in \mathscr{A}_Z$,
	\item \textbf{Linear:} $E(c_1 a_1 + c_2 a_2) = c_1 E(a_1) + c_2 E(a_2)$, 
	\item \textbf{Homogeneous:} $E(b a b^*) = b E(a) b^*, \qquad \forall b \in \mathscr{A}_S, a \in \mathscr{A}_Z$,
	\item \textbf{Unital:} $E(\op{1}) = \op{1}$. 
\end{enumerate}
It is straightforward to see that homogeneity and unitality together imply that $E$ is a projection: $E \circ E = E$. Thus, the algebraic map $E$ possesses all of the same properties as the map of the same name defined in the path integral language. In fact, one might argue that this algebraic formulation provides a more rigorous starting point for the modified holographic dictionary \eqref{eq:modAsymHolo}. Namely, the semiclassical gravity theory is dual to the theory of a subalgebra $\mathscr{A}_S \subset \mathscr{A}_Z$ contained inside the full large $N$ CFT, and the `filter' $E: \mathscr{A}_Z \rightarrow \mathscr{A}_S$ is simply a consistent algebraic projection from the full CFT to this subalgebra. 

Given an algebraic inclusion $B \subset A$, a map $E: A \rightarrow B$ satisfying the above considerations is called a \emph{conditional expectation}.\footnote{For a more complete introduction to conditional expectations and associated maps we refer the reader to Appendix \ref{app: QCon} and \cite{AliAhmad:2024saq,AliAhmad:2025oli}.} Remarkably, such a map possesses an alternative characterization which is \emph{intrinsic} to the algebra $B$. This characterization is categorical in nature and was introduced in the pioneering work of Longo and collaborators \cite{Longo:1994zza,Longo:1989tt,Bischoff:2014xea,DelVecchio:2017axj}. We will now give a very brief overview of this correspondence, including only those details which are relevant for the present discussion. For more technical details, we refer the reader to \cite{Bischoff:2014xea} and section~(3.3) of \cite{AliAhmad:2025oli}. The formal statement is as follows: there exists a $1-1$ correspondence between conditional expectations\footnote{Technically, the following discussion is oriented toward conditional expectations of finite index, however a generalizations to infinite index inclusions and even inclusions admitting only operator valued weights are also possible see \cite{AliAhmad:2025oli}.} onto the algebra $B$ and Q-systems inside the category $\text{End}(B)$. The category $\text{End}(B)$ has as its objects endomorphisms e.g. linear, unital maps from $B$ to itself which preserve its product and involution. Given two endomorphisms $\alpha_1, \alpha_2 \in \text{Obj}(\text{End}(B))$, the space of arrows $\text{Hom}_{\text{End}(B)}(\alpha_1,\alpha_2)$ consists of \emph{intertwiners} $w \in B$ such that $w \alpha_1(b) = \alpha_2(b) w$ for all $b \in B$. A Q-system is a triple $Q \equiv (\theta, x, w)$ consisting of an endomorphism $\theta \in \text{End}(B)$, an isometric intertwiner $w \in \text{Hom}_{\text{End}(B)}(\text{id},\theta)$ and an intertwiner $x \in \text{Hom}_{\text{End}(B)}(\theta, \theta \circ \theta)$ satisfying the compatibility conditions
\begin{enumerate}
	\item $w^* x = \theta(w^*) x = \op{1}$, and
	\item $x^2 = \theta(x) x$. 
\end{enumerate}

Perhaps the most interesting implication of the correspondence between conditional expectations and Q-systems is that it implies we can enlarge a given algebra using only data intrinsic to a smaller one. To be precise, from a conditional expectation $E: A \rightarrow B$, one can automatically construct a Q-system $Q_E = (\theta_E,x_E,w_E)$. Conversely, given a Q-system $Q = (\theta, x, w)$ within the category $\text{End}(B)$, one can also \emph{construct} an enlarged algebra $A_Q$ along with a conditional expectation $E_Q: A_Q \rightarrow B$ such that $Q_{E_Q} = Q$.\footnote{In Appendix \ref{app: incl}, we provide a discussion of this fact following from the Stinespring dilation theorem and a generalization of the Gram-Schmidt procedure for algebra valued inner products. See also section~\ref{subsec: SSandBU}.} Moreover, every inclusion of $B$ into a larger algebra which admits a conditional expectation can be obtained in this way. Thus, the collection of Q-systems associated with the algebra $B$ -- which can be defined in terms of the algebra $B$ intrinsically \emph{without} reference to the larger algebra -- classify the possible extensions of this algebra which admit conditional expectations. This is the crucial ingredient which will allow us to build our extended gravitational path integral. 

Each conditional expectation $E: \mathscr{A}_Z \rightarrow \mathscr{A}_{S}$ corresponds to a different Q-system $Q_E$ associated with the category $\text{End}(\mathscr{A}_S)$. These Q-systems classify the possible extensions of the `smooth' CFT algebra $\mathscr{A}_S$ into a `total' CFT algebra $\mathscr{A}_Z$. Let $\mathscr{A}_{\mathcal{Z}}$ be the algebra induced from the gravitational path integral. According to the modified holographic dictionary \eqref{eq:ModHol}, this algebra is dual to $\mathscr{A}_S$. Consequently, the Q-system $Q_E$ should be dual to a \emph{gravitational} Q-system $\mathcal{Q}_E$ associated with the category $\text{End}(\mathscr{A}_{\mathcal{Z}})$. Per the correspondence described above, the Q-system $\mathcal{Q}_E$ gives rise to an extended algebra $\mathscr{A}_{\textrm{ext.}}$ along with a conditional expectation $\mathcal{E}: \mathscr{A}_{\textrm{ext.}} \rightarrow \mathscr{A}_{\mathcal{Z}}$. In this regard, each possible choice of filter $E$ defines for us a different extended gravitational algebra. This is the first ingredient toward defining our extended gravitational path theory.

Using our correspondence between operator algebras and path integrals, the algebra $\mathscr{A}_{\textrm{ext.}}$ gives rise to an \emph{extended gravitational path integral} $\mathcal{Z}^{\textrm{ext.}}_{\mathcal{E}}$. Here, we have distinguished that our extended path integral depends upon the conditional expectation $\mathcal{E}$ which \emph{defines} the extended algebra. As discussed in Appendix \ref{app: AlgPI}, this construction moreover requires a choice of representation or equivalently a `vacuum' state on the algebra $\mathscr{A}_{\textrm{ext.}}$. A nice way to parameterize such a choice is to start with the vacuum state $\omega_0$ for the algebra $\mathscr{A}_{\mathcal{Z}}$ and compose it with a quantum channel $\mathcal{C}: \mathscr{A}_{\textrm{ext.}} \rightarrow \mathscr{A}_{\mathcal{Z}}$ resulting in a state $\omega \equiv \omega_0 \circ \mathcal{C}$ on $\mathscr{A}_{\textrm{ext.}}$. Let us suppose that the gravitational path integral $\mathcal{Z}$, which constructs expectation values with respect to $\omega_0$, can be expressed in the form
\beq \label{eq:OrigGPI}
	\mathcal{Z}(\textrm{BC}) = \int \mathscr{D} \Phi \; \mathcal{O}_{\textrm{BC}}[\Phi] e^{-S[\Phi]},
\eeq
where $\mathscr{D}\Phi$ includes, implicitly, a sum over all bulk geometries compatible with the boundary condition $\textrm{BC}$. In Appendix \ref{app: AlgPI}, we provide a path integral representation for a quantum channel; it corresponds to the inclusion of a new set of fields $\Phi_{\mathcal{C}}$ along with a new `interaction term' in the action $S_{\mathcal{C}}[\Phi,\Phi_{\mathcal{C}}]$ such that
\beq \label{eq:QChanPI}
	\mathcal{C}(\textrm{BC}) = \int \mathscr{D} \Phi_{\mathcal{C}} \; \mathcal{O}_{\textrm{BC}}[\Phi,\Phi_{\mathcal{C}}] \; e^{-S_{\mathcal{C}}[\Phi,\Phi_{\mathcal{C}}]}.
\eeq
As advertised, this map takes as input an `extended' boundary condition e.g. an insertion depending upon $(\Phi,\Phi_{\mathcal{C}})$ and outputs an expression depending only upon $\Phi$ which can be interpreted as the insertion associated with a boundary condition in the original theory. Combining \eqref{eq:OrigGPI} and \eqref{eq:QChanPI}, we can write down a new, extended path integral which computes expectation values in the state $\omega_0 \circ \mathcal{C}$:
\beq \label{eq:extGrav}
	\mathcal{Z}_{\mathcal{E},\mathcal{C}}^{\textrm{ext.}}(\textrm{BC}) = \mathcal{Z} \circ \mathcal{C}(\textrm{BC}) = \int \mathscr{D} \Phi \mathscr{D} \Phi_{\mathcal{E}} \; \mathcal{O}_{\textrm{BC}}[\Phi,\Phi_{\mathcal{E}}] \;  e^{-(S[\Phi] + S_{\mathcal{C}}[\Phi,\Phi_{\mathcal{E}}])}. 
\eeq 
We have relabeled the extended degrees of freedom $\Phi_{\mathcal{E}}$ to emphasize that they are determined through the conditional expectation $\mathcal{E}$, while the channel $\mathcal{C}$ determines the interaction $S_{\mathcal{C}}[\Phi,\Phi_{\mathcal{E}}]$. 

In summary, the extended gravitational path integral depends upon the choice of conditional expectation $\mathcal{E}$, which defines a set of extended degrees of freedom $\Phi_{\mathcal{E}}$, and the choice of channel $\mathcal{C}$, which tells how these degrees of freedom interact with the `smooth' gravitational degrees of freedom. This leads to the question, how should these objects be determined? As we will now describe, provided that $(\mathcal{E},\mathcal{C})$ is judiciously chosen, the inclusion of the new degrees of freedom it implicates and their associated interaction can be used to resolve the various inconsistencies described in section~\ref{sec: obst}. In this regard, we would like to interpret $\mathcal{Z}^{\textrm{ext.}}_{\mathcal{E},\mathcal{C}}$ as an intermediary between the naive semiclassical gravity theory $\mathcal{Z}$ and a complete non-perturbative quantum gravity theory. The physics described by $(\mathcal{E},\mathcal{C})$ serve as a mathematical stand-in for a genuine non-perturbative completion of $\mathcal{Z}$. The resulting theory is not sufficient to probe every fine grained detail of the non-perturbative gravity theory. However, the new degrees of freedom provide enough structure to cure some of the mathematical inconsistencies that $\mathcal{Z}$ suffers from. In this respect, the construction of $\mathcal{Z}^{\textrm{ext.}}_{\mathcal{E},\mathcal{C}}$ can be viewed in the spirit of renormalization -- a point which we turn to now. 

\subsection{Wormhole Renormalization and the Factorization Problem} \label{subsec: WHRenorm}

The first major advantage of the modified duality \eqref{eq:ModHol} is that it eliminates the tension between a non-factorizing gravitational path integral and a factorizing quantum mechanical path integral. The subtheory $Z^{\textrm{S}}$ is not a \emph{complete} quantum theory, and so it is perfectly consistent for this theory to fail to factorize. In this regard, the modified duality reinforces a common point of view that the failure of factorization for the semiclassical gravity theory is a consequence of its `incompleteness' \cite{Heidenreich:2021xpr,McNamara:2021cuo}. 

This idea was advanced in a very interesting way in \cite{Gesteau:2024gzf}. The main argument of that work is that the non-factorization of the semiclassical gravitational path integral should be understood in the same spirit as UV divergences which are encountered when computing bare quantities in an effective field theory. Namely, both are signals of the incompleteness of an effective description. To circumvent this problem in perturbative quantum field theory, one adds UV divergent counterterms to the effective action which cancel the bare divergences such that the renormalized correlation functions are finite. These counterterms can subsequently be calibrated such that the finite correlation functions match any data within the regime of validity of the effective theory. 

Likewise, the authors argue that it should be possible to implement a similar `wormhole renormalization' scheme to diagnose and cure the non-factorization of a semiclassical gravitational path integral. The extended gravitational path integral \eqref{eq:extGrav} should be interpreted in just this light. As we have addressed, the extended path integral is obtained by incorporating new fluctuating fields, $\Phi_{\mathcal{E}}$, which are subsequently integrated out against an interaction, $S_{\mathcal{C}}[\Phi,\Phi_{\mathcal{E}}]$. The imprint of this `averaging' on the original theory can be interpreted as the inclusion of `counterterms' which, if judiciously chosen, can be used to ensure factorization for the extended path integral $\mathcal{Z}^{\textrm{ext.}}_{\mathcal{E},\mathcal{C}}$.

In this sense, the factorization problem contributes the first `constraint' which we can place on $(\mathcal{E},\mathcal{C})$. In particular, the channel $\mathcal{C}$ should suffer from a non-factorization which is precisely counter to the non-factorization of $\mathcal{Z}$ such that their composition factorizes. 
The impact of including the interaction $S_{\mathcal{C}}[\Phi,\Phi_{\mathcal{E}}]$ can be interpreted as a source of UV information which is fed back into the semiclassical gravity theory to ensure its mathematical consistency. 

\begin{figure}[h!]
    \centering
    \begin{tcolorbox}[
        colback=white!10!white, 
        colframe=black, 
        boxrule=0.8pt, 
        arc=4pt, 
        sharp corners,
        halign=center,
        left=2mm, right=2mm, top=1mm, bottom=1mm, 
        width=0.8\linewidth 
    ]
        $\mathcal{C}$ is chosen such that $\mathcal{Z}^{\textrm{ext.}}_{\mathcal{E},\mathcal{C}} \equiv \mathcal{Z} \circ \mathcal{C}$ is a factorizing path integral
    \end{tcolorbox}
    \caption{To resolve the factorization problem, the channel $\mathcal{C}$ must include `counterterms' which cancel the contributions of connected wormholes in $\mathcal{Z}$.}
    \label{fig:factorizing-path-integral}
\end{figure}

It is worth noting that a rather explicit example of the above mechanism appears in low dimensional gravity. For example, in \cite{Blommaert:2021fob} a factorizing two-dimensional gravity theory is realized by incorporating correlated spacetime branes which carry a non-local interaction.\footnote{We thank Andreas Blommaert for bringing this paper to our attention.} This mechanism is closely related to the factorization realized by the inclusion of so-called `half wormholes' also discussed in e.g. \cite{Saad:2021rcu,Mukhametzhanov:2021hdi}. It has been proposed that these ideas have a dual avatar on the CFT side in terms of the subdivision of the CFT partition function into a sum of two terms; an `average' piece that captures the universal contribution of light CFT operators and a `deviation' piece that encodes a finer underlying microscopic structure \cite{Benjamin:2021ygh,DiUbaldo:2023qli}. This interpretation of the CFT partition function maps rather directly onto the `smooth' vs. `erratic' decomposition advocated for in \cite{Liu:2025ikq}.

\subsection{Conditional Entropy and the Information Problem} \label{subsec: condEnt}

As we have addressed in section~\ref{subsec: GPI-QI}, the factorization problem and the information problem are very intimately related. As we will now show, so too are their resolutions. Indeed, whereas the obstruction to factorization was caused by a non-trivial contribution to the gravitational path integral from connected wormhole saddles, the failure of the gravitational replica trick to reproduce the desired Page curve can be traced back to the fact that non-factorization implies we must subtract the connected contribution of the path integral away in the computation of the Renyi entropy. In the previous subsection, we explained how the extended gravitational path integral incorporates new counterterms, defined through the channel $\mathcal{C}$, which can be used to cancel the non-factorization caused by $\mathcal{Z}_{\textrm{conn.}}$. We will now describe how, in the entropy computation, these counterterms \emph{add back} the would-be contribution of $\mathcal{Z}_{\textrm{conn.}}$ such that the gravitational von Neumann entropy is equivalent to \eqref{NaiveGPIEnt}, the quantity which reproduces the Page curve. 

To understand this point, it is again useful to transpose our analysis into an operator algebraic language. The analysis of section~\ref{subsec: GPI-QI} allowed us to conclude that the gravitational entropy of a state $\psi_R$ defined through the gravitational path integral $\mathcal{Z}$ is given by
\beq
	\mathcal{S}(\psi_R) = \lim_{n \rightarrow 1} \frac{1}{1-n} \bigg[\log\bigg(\mathcal{Z}(\bigsqcup_{i = 1}^n M, \textrm{BC}_{\Psi^{\otimes n},s_n^R})\bigg) - \log\bigg(\mathcal{Z}_{\textrm{conn.}}(\bigsqcup_{i = 1}^n M, \textrm{BC}_{\Psi^{\otimes n},s_n^R})\bigg)\bigg].
\eeq
Crucially, this entropy is computed with respect to the gravitational algebra $\mathscr{A}_{\mathcal{Z}}$. Using the channel $\mathcal{C}: \mathscr{A}_{\textrm{ext.}} \rightarrow \mathscr{A}_{\mathcal{Z}}$ we can extend the state $\psi_R$ on $\mathscr{A}_{\mathcal{Z}}$ to a state $\psi_R \circ \mathcal{C}$ on the \emph{extended} gravitational algebra $\mathscr{A}_{\textrm{ext.}}$. In Appendix~\ref{app: condEnt}, we review a computation originally presented in \cite{AliAhmad:2024saq} which allows for the entropy of the state $\psi_R \circ \mathcal{C}$ to be decomposed as
\beq
	\mathcal{S}(\psi_R \circ \mathcal{C}) = \mathcal{S}(\psi_R) + \mathcal{S}(\mathcal{C}).
\eeq	
The latter term is called the \emph{conditional entropy} of $\mathcal{C}$. Provided the channel $\mathcal{C}$ is chosen such that
\beq
	\mathcal{S}(\mathcal{C}) = \lim_{n \rightarrow 1} \frac{1}{1-n} \log\bigg(\mathcal{Z}_{\textrm{conn.}}(\bigsqcup_{i = 1}^n M, \textrm{BC}_{\Psi^{\otimes n},s_n^R})\bigg) + \dots,
\eeq
we can conclude that
\beq
	\mathcal{S}(\psi_R \circ \mathcal{C}) = \lim_{n \rightarrow 1} \frac{1}{1-n} \log\bigg(\mathcal{Z}(\bigsqcup_{i = 1}^n M, \textrm{BC}_{\Psi^{\otimes n},s_n^R})\bigg) + \dots . 
\eeq
Here, $+ \dots$ refers to contributions to the conditional entropy which depend only on the extended degrees of freedom and are therefore independent of the original system. 

We therefore arrive at our second `constraint' on the channel $\mathcal{C}$: 
\begin{figure}[H]
    \centering
    \begin{tcolorbox}[
        colback=white!10!white, 
        colframe=black, 
        boxrule=0.8pt, 
        arc=4pt, 
        sharp corners,
        halign=center,
        left=2mm, right=2mm, top=1mm, bottom=1mm, 
        width=0.8\linewidth 
    ]
        The conditional entropy of $\mathcal{C}$ is given by \\ $\mathcal{S}(\mathcal{C}) = \lim_{n \rightarrow 1} \frac{1}{1-n} \log\bigg(\mathcal{Z}_{\textrm{conn.}}(\bigsqcup_{i = 1}^n M, \textrm{BC}_{\Psi^{\otimes n},s_n^R})\bigg) + \dots$
    \end{tcolorbox}
    \caption{To resolve the information problem, the quantum conditional entropy associated with the map $\mathcal{C}$ must contain the would be entropic contribution of the replica wormholes.}
    \label{fig:page-curve-path-integral}
\end{figure}

\noindent This ensures that the von Neumann entropy of the state $\psi_R \circ \mathcal{C}$ on the extended gravitational algebra $\mathscr{A}_{\textrm{ext.}}$ reproduces the Page curve. As we have alluded to above, this constraint is quite consistent with the constraint enshrined in Figure \ref{fig:factorizing-path-integral}. Namely, if the channel contributes counterterms which cancel non-factorization, the information theoretic contribution of these counterterms will \emph{restore} the connected component of the gravitational path integral in an extended gravitational replica trick computation. 

\subsection{Superselection Sectors and the Closed Universe Problem} \label{subsec: SSandBU}

Finally, let us consider the question of whether our extended gravitational path integral can accommodate the emergence of non-trivial closed universes. Once again, the answer to this question reveals a satisfying compatibility between our three problems and their single unified resolution. Rather than imposing a constraint on the channel $\mathcal{C}$, the resolution of this problem imposes a structural constraint on the conditional expectation $\mathcal{E}$ used to perform the algebraic extension of $\mathscr{A}_{\mathcal{Z}}$ to $\mathscr{A}_{\textrm{ext.}}$.  Recall that this conditional expectation is dual, in the holographic sense, to the conditional expectation $E: \mathscr{A}_Z \rightarrow \mathscr{A}_S$ on the CFT side. In this regard, the following discussion may alternatively be interpreted as a set of conditions for determining the algebra $\mathscr{A}_Z$, as an algebraic extension of $\mathscr{A}_S$. As we will discuss in section~\ref{sec: discuss}, this can be viewed as a \emph{definition} of a modified large $N$ limit.

The algebra $\mathscr{A}_{\textrm{ext.}}$ is obtained from $\mathscr{A}_{\mathcal{Z}}$ along with the Q-system $Q = (\theta,x ,w)$ in $\text{End}(\mathscr{A}_{\mathcal{Z}})$ which is induced from the holographically dual conditional expectation $E: \mathscr{A}_{Z} \rightarrow \mathscr{A}_S$. Let us now be a bit more detailed about this construction. Assume that $i: \mathscr{A}_{\mathcal{Z}} \hookrightarrow \mathscr{A}_{\textrm{ext.}}$ is the putative inclusion associated with our extended algebra. Then, it is shown in e.g. \cite{Bischoff:2014xea} that $\mathscr{A}_{\textrm{ext.}}$ can be defined simply as the algebraic union of $i(\mathscr{A}_{\mathcal{Z}})$ and the symbol $v$ which is defined by the following algebraic relations:
\beq \label{eq:QExtRel}
	v i(a) = i \circ \theta(a) v, \qquad v^2 = i(x) v, \qquad v^* = i(w^* x^*) v, \;\; \qquad \forall a \in \mathscr{A}_{\mathcal{Z}}. 
\eeq
We refer the reader to section~(3.3) of \cite{AliAhmad:2025oli} for a complete, pedagogical demonstration that $\mathscr{A}_{\textrm{ext.}} \equiv \mathscr{A}_{\mathcal{Z}} \vee \{v\}$ is indeed an associative $*$-algebra containing $\mathscr{A}_{\mathcal{Z}}$ as a consistent subalgebra. 

The physical interpretation of $\mathscr{A}_{\textrm{ext.}}$ is aided by a bit of unpacking. The first ingredient appearing in the Q-system, $\theta \in \text{End}(\mathscr{A}_{\mathcal{Z}})$, is an endomorphism of the algebra $\mathscr{A}_{\mathcal{Z}}$. Such a map may be viewed as a (possibly unfaithful) representation of $\mathscr{A}_{\mathcal{Z}}$. We say that a pair of endomorphisms, $\theta_1,\theta_2 \in \text{End}(\mathscr{A}_{\mathcal{Z}})$, are \emph{unitarily equivalent} if there exists a unitary $u \in \text{Hom}_{\text{End}(\mathscr{A}_{\mathcal{Z}})}(\theta_1,\theta_2)$ e.g. a unitary $u \in \mathscr{A}_{\mathcal{Z}}$ such that
\beq
	u \theta_1(a) = \theta_2(a) u \iff \theta_1(a) = u^{-1} \theta_2(a) u, \qquad \forall a \in \mathscr{A}_{\mathcal{Z}}. 
\eeq
The equivalence classes of endomorphisms induced by unitary equivalence are called \emph{sectors} -- we will denote by $[\theta]$ the sector associated with $\theta \in \text{End}(\mathscr{A}_{\mathcal{Z}})$. Related, an endomophism $\alpha$ is called irreducible if $\text{Hom}_{\text{End}(\mathscr{A}_{\mathcal{Z}})}(\alpha,\alpha) = \mathbb{C}$. 

Generalizing a familiar result from representation theory, the endomorphism $\theta$ may be decomposed into a direct sum of \emph{irreducible} endomorphisms, or more accurately any sector can be reduced to a direct sum of irreducible sectors:
\beq
	[\theta] = \bigoplus_{\gamma \prec \theta} [\gamma].
\eeq
Here, $\gamma \prec \theta$ implies the existence of an isometric, but not necessarily unitary, intertwiner $s \in \text{Hom}_{\text{End}(\mathscr{A}_{\mathcal{Z}})}(\gamma,\theta)$ such that $s^* s = \op{1}$. For each $\gamma \prec \theta$ contained in the irreducible sector decomposition of $\theta$ we therefore obtain an isometry $w_{\gamma} \in \text{Hom}_{\text{End}(\mathscr{A}_{\mathcal{Z}})}(\gamma,\theta)$. In fact, the collection $\{w_{\gamma}\}_{\gamma \prec \theta}$ can be interpreted as an irreducible decomposition of the intertwining element $w \in \text{Hom}_{\text{End}(\mathscr{A}_{\mathcal{Z}})}(\text{id},\theta)$ which appears in the Q-system. 

Given the irreducible decomposition of $\theta$, we can define a collection of operators
\beq
	\lambda_{\gamma} \equiv i(w_{\gamma}^*) v.
\eeq
From eqn. \eqref{eq:QExtRel}, the defining relations of the extension $\mathscr{A}_{\textrm{ext.}}$, we deduce that
\beq
	\lambda_{\gamma} i(a) = i \circ \gamma(x) \lambda_{\gamma}, \qquad \forall a \in \mathscr{A}_{\mathcal{Z}}. 
\eeq
For this reason, the operators $\lambda_{\gamma}$ are often referred to as \emph{charged intertwiners} since they map $a \in \mathscr{A}_{\mathcal{Z}}$ from the trivial representation into the representation associated with the irreducible sector $[\gamma]$. A general operator $\mathfrak{X} \in \mathscr{A}_{\textrm{ext.}}$ can be written in the form
\beq
	\mathfrak{X} = \sum_{\gamma \prec \theta} i(a_{\gamma}) \lambda_{\gamma},
\eeq
where $a_{\gamma} \in \mathscr{A}_{\mathcal{Z}}$ is a possibly different element for each sector $[\gamma]$. In words, the extended algebra can be interpreted as $\mathscr{A}_{\mathcal{Z}}$ `fibered' over its collection of irreducible sectors along a collection of new operators $\{\lambda_{\gamma}\}_{\gamma \prec \theta}$ which explicitly intertwine between the copies of the algebra located at each sector. 

This brings us to our final `constraint' which is now placed on the conditional expectation $\mathcal{E}$: \\
\begin{figure}[H]
    \centering
    \begin{tcolorbox}[
        colback=white!10!white, 
        colframe=black, 
        boxrule=0.8pt, 
        arc=4pt, 
        sharp corners,
        halign=center,
        left=2mm, right=2mm, top=1mm, bottom=1mm, 
        width=0.8\linewidth 
    ]
        The irreducible sectors $[\gamma]$ of the Q-system associated with $\mathcal{E}$ construct closed universe states
    \end{tcolorbox}
    \caption{To resolve the closed universe problem, the extended degrees of freedom induced from the conditional expectation $\mathcal{E}$ should furnish creation and annihilation operators for \emph{semiclassical} gravitational superselection sectors.}
    \label{fig:BU-sectors}
\end{figure}

\noindent If this is true, the algebra formed by the charged intertwiners $\{\lambda_{\gamma}\}_{\gamma \in \theta}$ can be regarded as encoding the non-trivial degrees of freedom associated with creating and annihilating closed universes which are entangled with the original system $\mathscr{A}_{\mathcal{Z}}$. Again, this constraint is by its very nature consistent with the previous two. As we have addressed in section~\ref{subsec: WHRenorm}, it is the role of the conditional expectation $\mathcal{E}$ to determine the new degrees of freedom and the role of the channel $\mathcal{C}$ to determine how these new degrees of freedom interact with the original ones. If the channel restores factorization and adds back the contribution of connected wormhole topologies to the gravitational entropy, then it stands to reason that the degrees of freedom associated with the extension must have an interpretation as describing a non-trivial gravitational system which the original theory becomes entangled with.

A few comments are in order relating our construction to the $\alpha$-parameters of Coleman \cite{Coleman:1988cy}, Giddings and Strominger \cite{Giddings:1988wv}, and Marolf and Maxfield \cite{Marolf:2020rpm}. Our $[\gamma]$-sectors are induced from the conditional expectation $\mathcal{E}$ which contains data about all of the non-factorization pathologies of $\mathcal{Z}$. By contrast, the $\alpha$-sectors of Marolf and Maxfield only diagnose non-factorization at the partition function level.\footnote{We thank Zhencheng Wang for helpful discussion on this point, which will be explored in detail in his forthcoming work with Jake McNamara \cite{McNamaraWang}.} At the same time, the role played by the sectors $[\gamma]$ in realizing factorization is fundamentally different from the $\alpha$-sectors. In our case, factorization is achieved by extending the theory, whereas in the Marolf-Maxfield analysis, one realizes factorization by restricting to any single $\alpha$-sector. Nevertheless, there are some structural similarities between the $[\gamma]$ and $\alpha$ sectors and it would be interesting to investigate to explore whether they are physically related in any way. 

\subsection{Existence and Uniqueness}

As advertised, what we have done in the preceding sections is set up a bootstrapping problem. Given the semiclassical gravitational path integral, $\mathcal{Z}$ and its corresponding algebra $\mathscr{A}_{\mathcal{Z}}$ we seek
\begin{enumerate}
\renewcommand{\labelenumi}{\alph{enumi}.}
	\item An extended gravitational algebra $\mathscr{A}_{\textrm{ext.}}$ defined by a Q-system associated with conditional expectation $\mathcal{E}: \mathscr{A}_{\textrm{ext.}} \rightarrow \mathscr{A}_{\mathcal{Z}}$, and
	\item A quantum channel $\mathcal{C}: \mathscr{A}_{\textrm{ext.}} \rightarrow \mathscr{A}_{\mathcal{Z}}$
\end{enumerate}
such that 
\begin{enumerate}
	\item The resulting extended path integral $\mathcal{Z}^{\textrm{ext.}}_{\mathcal{E},\mathcal{C}} = \mathcal{Z} \circ \mathcal{C}$ factorizes,
	\item The conditional entropy of the channel $\mathcal{C}$ includes a contribution equal to the entropy of the connected component of $\mathcal{Z}$, and
	\item The superselection sectors identified by $\mathcal{E}$ furnish charged intertwiners that construct closed universe states.
\end{enumerate}

Given such a problem, the natural questions to ask are whether it admits any solutions and, in the event that it does, whether the solutions are unique. In forthcoming work, we plan to investigate these questions directly by analyzing the possible Q-system extensions of the large N single trace algebra, and considering our program in the context of $3d$ gravity.\footnote{We should note that a related idea has been proposed in \cite{AliAhmad:2024saq} toward formulating the quantum extremal surface prescription purely from the boundary point of view.} For now, we can make a few preliminary comments. The first two constraints can be interpreted as defining a general information theoretic problem. The first is related to the notion of wormhole renormalization introduced in \cite{Gesteau:2024gzf}. Under the considerations of that note, we expect this problem should admit solutions. The second can be formulated in the following way: Fix an algebra $\mathscr{B}$ and let $\psi^{(n)}$ be a generally non-factorizing state on $\mathscr{B}^{\otimes n}$. Then, we seek an inclusion $\mathscr{B} \subset \mathscr{B}_{\textrm{ext.}}$, a quantum channel $\Upsilon: \mathscr{B}_{\textrm{ext.}} \rightarrow \mathscr{B}$ and a state $\psi$ on $\mathscr{B}$ such that
\beq \label{eq:repQIProblem}
	(\psi \circ \Upsilon)^{\otimes n}(s_n^{\mathscr{B}_{\textrm{ext.}}}) = \psi^{(n)}(s_n^{\mathscr{B}}) + ... \; . 
\eeq 
Here, $s_n^{\mathscr{B}}$ and $s_n^{\mathscr{B}_{\textrm{ext.}}}$ are the swap operators for $\mathscr{B}$ and $\mathscr{B}_{\textrm{ext.}}$, respectively. One can interpret $\psi^{(n)}$ as the semiclassical gravitational path integral with $n$-fold replica boundary conditions. Conversely, $(\psi \circ \Upsilon)^{\otimes n}$ should be interpreted as the factorizing extended path integral with appropriately extended replica boundary conditions. In this sense, we see that the first two constraints are in fact closely related. Assuming we can wormhole renormalize the gravitational path integral to obtain a factorizing extended version, we expect this wormhole renormalized path integral should also satisfy eqn. \eqref{eq:repQIProblem}. 

This leaves only the third constraint. To begin, we can see that this constraint is closely related to the first two by considering the entropy of a state on the extended algebra. Provided the first two conditions are met, we find\footnote{The formula \eqref{ext ent} should be compared with the realization of the generalized entropy as part of a von Neumann entropy by implementation of the gravitational constraints e.g. as in \cite{Klinger:2026tws}. In that case we extend the subregion algebra of the QFT $M \subset B(\mathscr{H})$ to the crossed product $M \subset M \times_{\alpha} G \subset B(\mathscr{H} \otimes L^2(G))$, where $G$ is the group of large diffeomorphisms. Given a wavefunction $\ket{\xi} \in L^2(G)$ we can extend a state $\ket{\varphi} \in \mathscr{H}$ to a `classical-quantum' state $\ket{\tilde{\varphi}_{\xi}} \equiv \ket{\varphi,\xi} \in \mathscr{H} \otimes L^2(G)$. Such a state has an entropy
\beq
	S(\tilde{\varphi}_{\xi}) = S_{\textrm{gen.}}(\varphi) + S_{\textrm{vN}}(\xi). 
\eeq
Eqn. \eqref{ext ent} is a generalization of this computation, as explained in \cite{AliAhmad:2024saq}.}
\beq \label{ext ent}
	\mathcal{S}(\psi_R \circ \mathcal{C}) = \mathcal{H}(\psi_R) + \mathcal{S}_{\textrm{ext.}}(\mathcal{C})
\eeq
where $\mathcal{H}(\psi_R)$ is the would-be entropy that arises from the naive gravitational replica trick \eqref{NaiveGPIEnt}, and $\mathcal{S}_{\textrm{ext.}}(\mathcal{C})$ is the part of the extended entropy that depends only on the channel $\mathcal{C}$. As we have already discussed, in the case that $R$ coincides with the exterior of a black hole, this realizes a true von Neumann entropy reproducing the Page curve. We can also apply this set up to the entropy of a closed universe. In that case $\mathcal{H}(\psi_R) = 0$ -- which is one source of the claim that the closed universe Hilbert space is one-dimensional. However, this claim is questionable for the same reason that the resolution to the information problem is questionable, namely $\mathcal{H}(\psi_R)$ is \emph{not} the von Neumann entropy of any single state. On the other hand, the quantity \eqref{ext ent} \emph{is} the von Neumann entropy of a single state and we now find that
\beq
	\mathcal{S}(\psi_R \circ \mathcal{C}) = \mathcal{S}_{\textrm{ext.}}(\mathcal{C}) > 0
\eeq	
for the closed universe case. Consequently, the same modification that resolves factorization and reproduces the Page curve also automatically tells us that the entropy of a closed universe is non-zero. Moreover, the degrees of freedom which furnish the closed universe algebra are precisely the new operators appended to $\mathscr{A}_{\mathcal{Z}}$, e.g. the charged intertwiners. 

Unlike the first two constraints, the third introduces a level of interpretation which is related to the desire that the auxiliary degrees of freedom used to satisfy the first two constraints be `physical' in a satisfactory way. Put differently, the first two constraints admit many non-unique solutions employing a variety of different `wormhole renormalization schemes', each of which implicates a different set of auxiliary degrees of freedom. From this point of view, the third `constraint' is perhaps better viewed as an organizing principle for interpreting different candidate extended algebras. As we will discuss in the next section, this can be compared to the problem of defining a modified large $N$ limit on the CFT side, since the extended gravitational algebra can be identified as dual to the complete algebra of operators with a good large $N$ limit. 

\section{A Synthesis} \label{sec: discuss}

In this note, we reviewed three mathematical inconsistencies that plague semiclassical gravity: the Factorization Problem, the Information Problem, and the Closed Universe Problem. We then argued that all three of these inconsistencies could be resolved via a modification of the holographic dictionary. In accord with the discussion of \cite{Liu:2025ikq}, a semiclassical gravity theory cannot be dual to the full large $N$ CFT due to erratic contributions in the latter. Thus, the semiclassical gravity theory must be dual to a `filtered' version of the CFT which restricts to the degrees of freedom therein which depend `smoothly' on $N$. Building upon this observation, we proposed that some finite $N$ data encoded in the filtering map can be used to define an \emph{extended} gravitational theory. Algebraically, the extended gravitational theory appends to the semiclassical algebra of observables a collection of new operators, $\{\lambda_{\gamma}\}$, which intertwine between different superselection sectors of the former. From the path integral point of view, the extended gravitational theory is of the form
\beq \label{eq:FullGravDisc}
	\mathcal{Z}^{\textrm{ext.}}_{\mathcal{E},\mathcal{C}}(\textrm{BC}) = \int \mathscr{D} \Phi \mathscr{D} \Phi_{\mathcal{E}} \; \mathcal{O}_{\textrm{BC}}[\Phi,\Phi_{\mathcal{E}}] \; e^{-(S[\Phi] + S_{\mathcal{C}}[\Phi,\Phi_{\mathcal{E}}])}.
\eeq
Here, $\Phi_{\mathcal{E}}$ are new degrees of freedom which may be interpreted as the path integral version of the intertwining operators $\lambda_{\gamma}$, and $S_{\mathcal{C}}[\Phi,\Phi_{\mathcal{E}}]$ is a new interaction between these degrees of freedom and the smooth gravity degrees of freedom. 

\begin{figure}[ht]
\centering
\begin{tikzpicture}[
    every node/.style={align=center},
    main/.style={font=\bfseries},
    eqbox/.style={
        draw,
        rounded corners,
        font=\small,
        inner sep=4pt
    },
    arrow/.style={-{Latex[length=3mm,width=2mm]}, thick}
]

\node[main] (H) at (90:4) {Modified Semiclassical \\ Holographic Principle};
\node[main] (G) at (210:4) {Gravitational\\ Path Integral};
\node[main] (Q) at (330:4) {Quantum\\ Information};

\coordinate (HGmid) at ($(H)!0.5!(G)$);
\coordinate (GQmid) at ($(G)!0.5!(Q)$);
\coordinate (HQmid) at ($(H)!0.5!(Q)$);

\node[eqbox] (F) at (HGmid)
{$\mathcal{Z}^{\textrm{ext.}}_{\mathcal{E},\mathcal{C}} = \mathcal{Z} \circ \mathcal{C}$};

\node[eqbox] (I) at (GQmid)
{$\mathcal{S}(\mathcal{C}) = \mathcal{S}_{\textrm{conn.}} + \dots$};

\node[eqbox] (L) at (HQmid)
{$\mathscr{A}_{\textrm{ext.}} = \sum_{\gamma \prec \theta}
  i(\mathscr{A}_{\mathcal{Z}}^{(\gamma)}) \lambda_{\gamma}$};

\draw[arrow] (H) -- (F);
\draw[arrow] (H) -- (L);

\draw[arrow] (G) -- (F);
\draw[arrow] (G) -- (I);

\draw[arrow] (Q) -- (I);
\draw[arrow] (Q) -- (L);

\end{tikzpicture}
\caption{Conditions required to recover a consistent mathematical theory with the modified holographic dictionary.}
\label{fig:ModQGTriangle}
\end{figure}

The interaction $S_{\mathcal{C}}[\Phi,\Phi_{\mathcal{E}}]$ is not a priori uniquely determined by the algebraic extension. As such, we have proposed a sequence of `consistency conditions' which can be used to restrict the possible choices. In particular, we noted that if $S_{\mathcal{C}}[\Phi,\Phi_{\mathcal{E}}]$ possesses a non-factorization problem which is appropriately `equal and opposite' to that of the semiclassical gravitational path integral, then the combined theory defined by \eqref{eq:FullGravDisc} will no longer suffer from non-factorization. This can be viewed as a form of `wormhole renormalization' \cite{Gesteau:2024gzf} in which the action $S_{\mathcal{C}}[\Phi,\Phi_{\mathcal{E}}]$ is interpreted as containing counterterms added to the semiclassical gravity theory to cancel its non-factorization. From a quantum information theoretic point of view, this also suggests that the interaction $S_{\mathcal{C}}[\Phi,\Phi_{\mathcal{E}}]$ should provide a contribution to the entropy which is `equal and opposite' to the would-be contribution of replica wormholes in the semiclassical gravitational replica trick. Consequently, the von Neumann entropy of states defined by the \emph{extended} gravitational path integral can be shown to satisfy a Page curve. 

The final constraint on the extended theory pertains to the identity of the extended degrees of freedom. In general, these degrees of freedom can be shown to relate to superselection sectors in the semiclassical theory. However, it is still somewhat ambiguous how these degrees of freedom should be interpreted physically. To conclude, we would therefore like to describe how our abstract mathematical construction resounds the features of many proposed approaches to understanding semiclassical quantum gravity. We find it encouraging that these different points of view might be unified, at least quantitatively, under a single umbrella. 

\subsection{No Global Symmetries and Background Independence}

The extension of the semiclassical algebra $\mathscr{A}_{\mathcal{Z}}$ to $\mathscr{A}_{\textrm{ext.}}$ underscores a correspondence between (a) the notion that quantum gravity should admit no global symmetries and (b) that quantum gravity should satisfy some form of background independence. To the first point, the Q-system used to extend the semiclassical gravity theory also describes a generalized symmetry of the algebra $\mathscr{A}_{\mathcal{Z}}$ \cite{AliAhmad:2025bnd}. Forming the extension $\mathscr{A}_{\textrm{ext.}}$ can be viewed as a generalized gauging of this symmetry, promoting it from global to local via the inclusion of new operations, $\{\lambda_{\gamma}\}_{\gamma \prec \theta}$, internal to the algebra $\mathscr{A}_{\textrm{ext.}}$ that implement the symmetry. At the same time, the various sectors $[\gamma]$ associated with this symmetry correspond to different concrete representations of $\mathscr{A}_{\mathcal{Z}}$. Employing the point of view described in \cite{Klinger:2025tvg}, these representations constitute different backgrounds for the semiclassical gravity theory. Including the new operators $\lambda_{\gamma}$ that intertwine between different sectors then has the interpretation of imposing a form of background independence. 

\subsection{Defining the Large N Limit}

In \cite{Liu:2025cml}, it is argued that the inclusion of the `smooth' subtheory into the full dual theory can be understood as the inclusion of the algebra of large $N$ single trace operators into the full algebra of operators with a satisfactory large $N$ limit. However, it is ambiguous how this limit should be defined \cite{Liu:2025cml,Kudler-Flam:2025cki}. Within our construction, the conditional expectation $E$, or equivalently its associated Q-system, can be regarded as \emph{defining} different possible large $N$ limits e.g. extensions of the `simple' single trace algebra. On the bulk side, the inclusion of the semiclassical gravity theory into the extended gravity theory should be compared to the inclusion of the causal wedge into an algebra that also includes complex, non-local operators. The form of these operators are in turn determined by our chosen large $N$ limit. From this point of view, the consistency conditions we have proposed can be viewed as instructions for identifying the physically relevant large $N$ limit from within the space of all possible choices. 

\subsection{Ensemble Averaging and Closed Universe Cosmologies}

From the point of view described in the previous paragraph, the nontriviality of the inclusion $\mathscr{A}_{\mathcal{Z}} \subset \mathscr{A}_{\textrm{ext.}}$ implies the existence of a gravitating system which is entangled with the `simple' bulk operators contained in $\mathscr{A}_{\mathcal{Z}}$. In the context of the AdS/CFT correspondence, this can be viewed as a gluing of two AdS bulk spacetimes across a shared closed universe:

\begin{figure}[H]
\centering
\begin{tikzpicture}[
    scale=1,
    every node/.style={font=\small},
    bulk/.style={draw, thick, rounded corners=18pt},
    universe/.style={draw, thick, circle},
    entangle/.style={red, thick, dash pattern=on 2.5pt off 2pt}
]

\node[bulk, minimum width=3.6cm, minimum height=2.6cm] (L) at (-4,0) {};
\node at (-4,0) {$\mathscr{A}^{(L)}_{\mathcal{Z}}$};

\node[bulk, minimum width=3.6cm, minimum height=2.6cm] (R) at (4,0) {};
\node at (4,0) {$\mathscr{A}^{(R)}_{\mathcal{Z}}$};

\node[universe, minimum size=2.2cm] (C) at (0,0) {};
\node at (0,0) {$\{\lambda_{\gamma}\}_{\gamma \prec \theta}$};

\foreach \y in {0.6,0.2,-0.2,-0.6} {
    \draw[entangle]
        ($(L.east)+(-0.12,\y)$) -- ($(C.west)+(0.12,\y)$);
    \draw[entangle]
        ($(R.west)+(0.12,\y)$) -- ($(C.east)+(-0.12,\y)$);
}

\end{tikzpicture}
\caption{The extended gravitational algebra can be interpreted as a gluing of two (smooth) AdS bulk algebras with the set of auxillary operators $\{\lambda_{\gamma}\}_{\gamma \prec \theta}$ specified by the conditional expectation $\mathcal{E}$.}
\end{figure}

\noindent This picture resonates quite nicely with a proposal\footnote{See also \cite{Cooper:2018cmb,VanRaamsdonk:2021qgv,Antonini:2022opp} for some closely related work.} of Van Raamsdonk \cite{VanRaamsdonk:2020tlr} for realizing the closed universe cosmologies of Maldecena and Maoz \cite{Maldacena:2004rf} within the AdS/CFT correspondence. The construction of \cite{VanRaamsdonk:2020tlr} realizes a path integral description analogous to eqn. \eqref{eq:FullGravDisc}, where in our case the `auxiliary' degrees of freedom are those defined by the conditional expectation $\mathcal{E}$. One may interpret the resulting path integral as an ensemble of theories parameterized by these degrees of freedom and with probability distribution determined by the interaction $S_{\mathcal{C}}[\Phi,\Phi_{\mathcal{E}}]$. This notion of gluing may be given a rigorous interpretation as a form of gauging \cite{Torres:2025jcb}, and thereby connected explicitly to the first point in this discussion. We plan to explore this construction in detail in future work.

\subsection{Observers and Observer Rules}

An alternative perspective on the emergence of non-trivial closed universe physics has been proposed in \cite{Harlow:2025pvj,Abdalla:2025gzn,Engelhardt:2025azi,Akers:2025ahe,Higginbotham:2025clp}. The idea is to add new propagating degrees of freedom into the theory which represent an `observer'. The inclusion of the observer comes with a set of modified rules for doing computations e.g. in the gravitational path integral. The observer rules can also be formalized in terms of a quantum to classical channel that describes its entanglement with the rest of the universe it inhabits \cite{Engelhardt:2025azi}. It is tempting to argue that the combined effect of extending the gravitational algebra $\mathscr{A}_{\mathcal{Z}} \mapsto \mathscr{A}_{\textrm{ext.}}$ and specifying a quantum channel $\mathcal{C}: \mathscr{A}_{\textrm{ext.}} \rightarrow \mathscr{A}_{\mathcal{Z}}$ could mathematically encode the same data as these observer rules \emph{without} the need to add an observer by hand. In particular, different choices of $(\mathcal{E},\mathcal{C})$ correspond to instantiating different observer degrees of freedom (encoded in $\mathcal{E}$) with different observer rules (encoded in $\mathcal{C}$). The physical interpretation in this case does is quite distinct since the degrees of freedom which are being appended to the naive semiclassical theory still have a fundamentally gravitational origin.

\appendix
\renewcommand{\theequation}{\thesection.\arabic{equation}}
\setcounter{equation}{0}

\section*{Acknowledgments}

We would like to thank Shadi Ali Ahmad, Vijay Balasubramanian, Jose Calderon-Infante, Luca Ciambelli, Elliott Gesteau, Temple He, Nima Lashkari, Hong Liu, Sean McBride, Daniel Murphy, Jeremy van der Heijden, Shreya Vardhan, Akash Vijay, Alejandro Vilar L\'{o}pez, Zhencheng Wang, and Tom Yildirim for helpful discussions. We are also appreciative to Andreas Blommaert and Alejandro Vilar L\'{o}pez for helpful comments leading to revisions of the draft. This work was supported by the Heising-Simons foundation ``Observable Signatures of Quantum Gravity" collaboration and the Walter Burke Institute for Theoretical Physics. This material is also based upon work supported by the U.S. Department of Energy, Office of Science, Office of High Energy Physics, under Award Number DE-SC0011632. 

\pagebreak

\section{Quantum Conditional Probability} \label{app: QCon}

In this appendix we review the features of quantum conditional probability which will be utilized in the main text. Specifically, we show that the existence of an operator valued weight between a pair of algebras allows us to construct a basis of operators which we used to define the extended gravitational path integral in section~\ref{sec: res}. Then, we show that the existence of generalized conditional expectations associated with algebraic inclusions allows for the factorization of the entropy into a sum of terms. We use this fact in section~\ref{subsec: condEnt} to show that the entanglement entropy satisfying the Page curve can be interpreted as the von Neumann entropy of a state defined by the extended gravitational path integral. 

\subsection{Preliminaries}

The theory of unital $C^*$ algebras and their states is sometimes called `noncommutative measure theory', since it generalizes the axioms of the classical measure theory of random variables and probability distributions. From this point of view, a $C^*$ algebra $A$ is a noncommutative measure space and its elements $a \in A$ are noncommutative random variables. The role of a probability measure on this `space' is played by an algebraic state, $\psi \in S(A)$, which is a positive, linear, normalized map $\psi: A \rightarrow \mathbb{C}$ which computes the expectation value of random variable $a \mapsto \psi(a)$. Moving forward it will also be necessary to consider non-normalizable generalizations of states called weights. A weight on $A$ is a map $\psi: A_+ \rightarrow \mathbb{R}_+$ which is additive and positively homogeneous. It is not required to have finite norm. We will denote the space of weights on $A$ by $P(A)$.

To each state we can associate a faithful Hilbert space representation $\pi_{\psi}: A \rightarrow B(L^2(A;\psi))$ via the GNS construction. Let $\mathfrak{n}_{\psi} \equiv \{a \in A \; | \; \psi(a^* a) < \infty\}$ denote the domain of $\psi$ and $\mathfrak{k}_{\psi} \equiv \{a \in A \; | \; \psi(a) = 0\}$ denote its kernel. Then, the quotient space $\mathfrak{n}_{\psi}/\mathfrak{k}_{\psi}$ is a preclosed inner product space with
\beq \label{eq:GSNInner}
	\langle \eta_{\psi}(a_1), \eta_{\psi}(a_2) \rangle_{\psi} \equiv \psi(a_1^* a_2).
\eeq
Here, $\eta_{\psi}: \mathfrak{n}_{\psi} \rightarrow \mathfrak{n}_{\psi}/\mathfrak{k}_{\psi}$ is the projection of the quotient. The completion of $\mathfrak{n}_{\psi}/\mathfrak{k}_{\psi}$ in \eqref{eq:GSNInner} defines the GNS Hilbert space of $A$ with respect to $\psi$, $L^2(A;\psi)$. This Hilbert space naturally obtains a representation
\beq
	\pi_{\psi}(a_1) \eta_{\psi}(a_2) \equiv \eta_{\psi}(a_1 a_2),
\eeq
and the state $\psi$ obtains a vector representative $\Omega_{\psi} \equiv \eta_{\psi}(\op{1}) \in L^2(A;\psi)$ such that
\beq
	\psi(a) = \langle \Omega_{\psi}, \pi_{\psi}(a) \Omega_{\psi} \rangle_{\psi}. 
\eeq

As a matter of fact, there exists a far reaching generalization of the GNS construction which allows us to make contact with the theory of noncommutative \emph{conditional} probability. Classical conditional probability is a theory of the interrelation between many, possibly correlated, stochastic systems each of which is described by a classical measure space. Accordingly, noncommutative or quantum conditional probability is a theory of the interrelation between many, possibly entangled and correlated, noncommutative stochastic systems each of which is described by a noncommutative measure space e.g. a unital $C^*$ algebra. The central object in conditional probability theory is a completely positive map between algebras $\alpha: A \rightarrow B$. Recall, such a map is called positive if it maps positive operators in $A$ to positive operators in $B$. It is completely positive if the extension $\alpha \otimes \text{id}_n: A \otimes M_n(\mathbb{C}) \rightarrow B \otimes M_n(\mathbb{C})$ is positive for all $n \in \mathbb{N}$. 

In some sense, completely positive maps are the most general class of algebraic maps which are compatible with the Hilbert space picture. This observation is formalized by Stinespring's theorem \cite{Stinespring:1955eig}. Let $\alpha: A \rightarrow B(\mathscr{H})$ be a completely positive map, with $\mathscr{H}$ a Hilbert space. Then, there exists a Hilbert space $L^2(A-\mathscr{H};\alpha)$, a representation $\pi_{\alpha}: A \rightarrow B(L^2(A-\mathscr{H};\psi)$, and a map $W_{\alpha}: \mathscr{H} \rightarrow L^2(A-\mathscr{H};\alpha)$ such that
\beq \label{eq:spatimp}
	\alpha(a) = W_{\alpha}^{\dagger} \pi_{\alpha}(a) W_{\alpha}. 
\eeq
Given eqn. \eqref{eq:spatimp}, we say that the map $\alpha$ is spatially implemented on the Hilbert space $L^2(A-\mathscr{H};\alpha)$ with spatial implementer $W_{\alpha}$. As we have alluded to, and as our notation is meant to suggest, the Stinespring theorem is a generalization of the GNS construction. Let $\mathfrak{n}_{\alpha}$ denote the domain of the map $\alpha$ and $\mathfrak{k}_{\alpha}$ its kernel. Then, the space $\mathfrak{n}_{\alpha}/ \mathfrak{k}_{\alpha} \otimes \mathscr{H}$ is a preclosed inner produce space with pre-inner product
\beq
	\langle \eta_{\alpha}(a_1) \otimes v_1, \eta_{\alpha}(a_2) \otimes v_2 \rangle_{\alpha} \equiv \langle v_1, \alpha(a_1^* a_2) v_2 \rangle_{\mathscr{H}},
\eeq
where $\eta_{\alpha}: \mathfrak{n}_{\alpha} \rightarrow \mathfrak{n}_{\alpha}/\mathfrak{k}_{\alpha}$ is the projection of the quotient and $\langle \cdot, \cdot \rangle_{\mathscr{H}}$ is the inner product of the Hilbert space $\mathscr{H}$. This Hilbert space naturally obtains a representation
\beq
	\pi_{\alpha}(a_1)\big( \eta_{\alpha}(a_2) \otimes v \big) \equiv \eta_{\alpha}(a_1 a_2) \otimes v,
\eeq
and the spatial implementer of $\alpha$ is given by
\beq
	W_{\alpha}(v) \equiv \eta_{\alpha}(\op{1}) \otimes v \implies W_{\alpha}^{\dagger}(\eta_{\alpha}(a) \otimes v) = \alpha(a)\big(v\big).  
\eeq

A state on an algebra $A$, or more generally a weight, is a special case of a completely positive map from the algebra $A$ to the algebra $\mathbb{C}$. It is easy to see that the Stinespring Hilbert space $L^2(A-\mathbb{C};\psi) \simeq L^2(A;\psi)$ and $W_{\psi}(z) = \Omega_{\psi} z$ for any $z \in \mathbb{C}$. 

The space of completely positive maps includes as subspaces many of the most important classes of maps in noncommutative conditional probability, as we will review now. Suppose that $\alpha: A \rightarrow B(\mathscr{H})$ is a completely positive map and $\beta: B \rightarrow B(\mathscr{H})$ is a faithful representation of an algebra $B$ which is generically not equal to $A$ or a subalgebra of $A$. Then, the map $\alpha$ may satisfy some or none of the following properties:
\begin{enumerate}
	\item \textbf{Unitality:}
	\beq
		\alpha(\op{1}) = \op{1} \iff W_{\alpha} \text{ is an isometry}. 
	\eeq
	\item \textbf{$B$-Preserving:}
	\beq
		\alpha(A) \subseteq \beta(B) \iff \gamma \equiv \beta^{-1} \circ \alpha: A \rightarrow B \text{ is a CP map}.
	\eeq
	\item \textbf{B-Homogeneity:} Suppose now that $i: B \hookrightarrow A$ is an algebraic inclusion
	\beq
		\alpha(i(b_1) \; a \; i(b_2)) = \beta(b_1) \alpha(a) \beta(b_2) \iff W_{\alpha} \beta(b) = \pi_{\alpha} \circ i(b) W_{\alpha}.
	\eeq	
	\item \textbf{Weight-Preserving:} Suppose again that $i: B \hookrightarrow A$, that $\alpha$ is $B$-Preserving
	\beq
		\exists \psi \in P(A) \text{ s.t. } \psi = \psi \circ i \circ \gamma. 
	\eeq
\end{enumerate}

A completely positive and unital map is called a quantum channel. If it is also $B$-preserving, $\gamma$ will define a quantum channel from $A$ to $B$. A completely positive, unital, $B$-preserving and state-preserving map is called a generalized conditional expectation. A completely positive, $B$-preserving, $B$-Homogeneous, weight-preserving map is called an operator valued weight. Finally, a completely positive, unital, $B$-preserving, $B$-homogeneous, weight-preserving map is called a conditional expectation. This classification is summarized in Figure \ref{fig:CPMaps}. We should emphasize that, due to the lack of unitality, an operator valued weight can only preserve a weight, while a generalized conditional expectation will generically preserve a state. 

In general, one might view a completely positive map as a way of coarse-graining quantum information. The combination of homogeneity and unitality tells us that a conditional expectation, $E: A \rightarrow B$ satisfies $E \circ i = \text{Id}_B$. In other words, the information which is coarse-grained under the conditional expectation is, in some sense, `orthogonal' to the subalgebra $B$. This may be viewed as a form of quantum error correction -- the existence of a conditional expectation implies that the subalgebra $B$ can be protected from whatever noise is applied by $E$. In this regard, generalized conditional expectations and operator valued weights can be viewed as forms of non-exact quantum error correction. In the case of the generalized conditional expectation, the failure of homogeneity implies that the subalgebra $B$ is shuffled around such that it may be recovered but only in a distorted way. In the case of the operator valued weight, the failure of unitality implies that the map is not spatially isometric. Nevertheless, as we shall now exhibit, operator valued weights, generalized conditional expectations and conditional expectations each provide access to different important aspects of noncommutative conditional probability theory.

\begin{figure}[h]
\centering
\renewcommand{\arraystretch}{1.3}
\begin{tabular}{lcccc}
\toprule
 & \textbf{Unital} & \textbf{$B$-Preserving} & \textbf{$B$-Homogeneous} & \textbf{Weight-Preserving} \\
\midrule
\textbf{QC}
 & $\checkmark$ & $\;$ & $\;$ & $\;$ \\

\textbf{GCE}
 & $\checkmark$ & $\checkmark$ & $\;$ & $\checkmark$ \\

\textbf{OVW}
 & $\;$ & $\checkmark$ & $\checkmark$ & $\checkmark$ \\

\textbf{CE}
 & $\checkmark$ & $\checkmark$ & $\checkmark$ & $\checkmark$ \\
\bottomrule
\end{tabular}
\caption{A taxonomy for completely positive maps.}
\label{fig:CPMaps}
\end{figure}

\subsection{Inclusions and Extensions} \label{app: incl}

Let $i: B \hookrightarrow A$ be an inclusion of $C^*$ algebras admitting an operator valued weight $T: A \rightarrow B$, e.g. $T$ is a completely positive map from $A$ to $B$ satisfying the additional property of homogeneity. Homogeneity implies that $T$ can be used to define a $B$-valued inner product on $A$:
\beq
	G_T: (a_1,a_2) \in A \times A \mapsto G_T(a_1,a_2) \equiv T(a_1^* a_2) \in B. 
\eeq
Included elements $i(b) \in A$ for $b \in B$ are treated like `scalars' with respect to this inner product:
\beq
	G_T(a_1 i(b), a_2) = b^* G_T(a_1,a_2), \qquad G_T(a_1, a_2 i(b)) = G_T(a_1,a_2) b. 
\eeq
Consequently, we can apply a $B$-valued generalization of the Gram-Schmidt orthogonalization procedure to obtain a `basis' $\{\lambda_i\}_{i \in \mathcal{I}} \subset A$ such that a general operator in $A$ can be expanded as
\beq
	a = \sum_{j \in \mathcal{I}} i(b_j) \lambda_j. 
\eeq

Combining the above with the standard output of Stinespring's theorem, we see that an operator valued weight naturally gives rise to the triple
\beq
	\bigg(\pi_T: A \rightarrow B(L^2(A-L^2(B;\psi_0);T)), \;\; W_T, \;\; \{\lambda_i\}_{i \in \mathcal{I}}\bigg),
\eeq
where $(\pi_T, W_T)$ is the canonical purification of $T$ and $\{\lambda_i\}_{i \in \mathcal{I}}$ is the Gram-Schmidt basis for the $B$-valued inner product that $T$ induces. In \cite{AliAhmad:2025oli}, it has been shown that the converse is also true. Given the operator algebra $B$ along with
\begin{enumerate}
	\item A Hilbert space representation $\pi: B \rightarrow B(\mathscr{H})$,
	\item An intertwining map $W: L^2(B;\psi_0) \rightarrow B(\mathscr{H})$, and
	\item A collection of operators $\{\Lambda_i\}_{i \in \mathcal{I}} \subset B(\mathscr{H})$ satisfying
	\beq
		W^{\dagger} \Lambda_i \Lambda_j^{\dagger} W \in \pi_{\psi_0}(B), \qquad \mathscr{H} = \text{span}\bigg(\sum_{i \in \mathcal{I}} \pi(b) \Lambda_i W(\Omega_{\psi_0})\bigg),
	\eeq
\end{enumerate}
we can \emph{construct} an extended algebra $A \equiv \pi(B) \vee \{\Lambda_i\}_{i \in \mathcal{I}}$ such that $T: a \in A \mapsto W^{\dagger} a W \in \pi_{\psi_0}(B)$ is an operator valued weight. The triple $(\pi, W, \{\Lambda_i\}_{i \in \mathcal{I}})$ is called a \emph{spatial Q-system}. The set of spatial Q-systems for the algebra $B$ classify its possible extensions into larger algebras admitting operator valued weights. 

	
\subsection{Conditional Entropy} \label{app: condEnt}

Let $i: B \hookrightarrow A$ be an inclusion of $C^*$, and let $\psi \in S(A)$ be a state such that $\psi_0 \equiv \psi \circ i \in S(B)$ is also a state. Then, by Petz duality, we obtain a generalized conditional expectation $i_{\psi}^{\dagger}: A \rightarrow B$. To be precise, $i_{\psi}^{\dagger}$ is the formal adjoint intertwining the KMS inner product of $A$ with respect to $\psi$ and of $B$ with respect to $\psi_0$. These inner products are of the form, 
\beq
	g^{\textrm{KMS}}_{\psi}(a_1,a_2) \equiv \psi(a_1^* \sigma^{\psi}_{-i/2}(a_2)),
\eeq
and thus
\beq
	g_{\psi}^{\textrm{KMS}}(a,i(b)) = g_{\psi_0}^{\textrm{KMS}}(i_{\psi}^{\dagger}(a),b). 
\eeq
Such a construction can be applied more generally to any completely positive map \cite{PetzDuality}, see e.g. \cite{AliAhmad:2024saq} for a discussion of this construction in relation to quantum error correction. 

Given a pair of generic completely positive maps $\alpha,\beta: A \rightarrow B(\mathscr{H})$, we say that $\alpha$ is differentiable with respect to $\beta$ if it is spatially implemented on the Stinespring Hilbert space $L^2(A-\mathscr{H};\beta)$. In \cite{BELAVKIN198649}, it is shown that this implies the existence of a positive, self-adjoint operator $\rho_{\alpha \mid \beta}: L^2(A-\mathscr{H};\beta) \rightarrow L^2(A-\mathscr{H};\beta)$ such that
\beq
	\alpha(a) = \bigg(\rho_{\alpha \mid \beta}^{1/2} W_{\beta}\bigg)^{\dagger} \pi_{\beta}(a) \bigg(\rho_{\alpha \mid \beta}^{1/2} W_{\beta}\bigg). 
\eeq
The operator $\rho_{\alpha \mid \beta}$ is called the spatial derivative of $\alpha$ with respect to $\beta$. 

As Stinespring's theorem can be interpreted as a generalization of the GNS construction, the spatial derivative $\rho_{\alpha \mid \beta}$ can be regarded as a generalization of the spatial derivative of states (weights). Recall that a state $\varphi \in S(A)$ is called differentiable with respect to a state $\psi \in S(A)$ if it possesses a vector representative in the GNS Hilbert space, $\Omega_{\varphi} \in L^2(A;\psi)$. Viewing the GNS Hilbert space $L^2(A;\psi)$ as a Stinespring Hilbert space for $\psi$ regarded as a completely positive map, we can see that the existence of such a vector representative implies that $\varphi$, also viewed as a completely positive map, is spatially implemented on $L^2(A;\psi)$. This implies the existence of an operator $\rho_{\varphi \mid \psi}^{1/2}$ such that
\beq
	\rho_{\varphi \mid \psi}^{1/2}\big(\Omega_{\psi}\big) = \Omega_{\varphi}. 
\eeq

If $\varphi$ is differentiable with respect to $\psi$ we can define the relative entropy between these two states as
\beq \label{eq:relent}
	S(\varphi \parallel \psi) \equiv \langle \Omega_{\varphi}, \log(\rho_{\varphi \mid \psi}) \Omega_{\varphi} \rangle_{\psi}. 
\eeq
It can be shown that $\rho_{\varphi \mid \psi}$ is equal to the analytic continuation of the Connes' cocycle derivative of $\varphi$ by $\psi$, which itself can be written in terms of the relative modular operator. In this way, the formula eqn. \eqref{eq:relent} reproduces the perhaps more well known expression
\beq
	S(\varphi \parallel \psi) = \langle \Omega_{\varphi}, \log(\Delta_{\varphi \mid \psi}) \Omega_{\varphi} \rangle_{\psi}. 
\eeq
In the event that $A$ admits a tracial weight $\tau$, the von Neumann entropy of $\varphi$ with respect to $\tau$ is defined to be
\beq \label{eq:vNEntWRTtrace}
	S_{\tau}(\varphi) \equiv -S(\varphi \parallel \tau). 
\eeq
It is easy to show that $\rho_{\varphi \mid \tau}$ defines the density operator of $\varphi$ with respect to $\tau$ e.g. $\varphi(a) = \tau(\rho_{\varphi \mid \tau} a)$. Thus, eqn. \eqref{eq:vNEntWRTtrace} reproduces the standard formula for the von Neumann entropy up to a (possibly infinite) state independent constant related to the normalization of the trace. 

It has been shown that the existence of generalized conditional expectations for generic inclusions combined with the definition of the spatial derivative for completely positive maps can be used to derive the following non-commutative factorization. Let $i: B \hookrightarrow A$ be an inclusion and $\varphi,\psi \in S(A)$ a pair of states which restrict to states $\varphi_0 \equiv \varphi \circ i, \psi_0 \equiv \psi \circ i \in S(B)$. Then, if $\varphi$ is differentiable with respect to $\psi$ and $\varphi_0$ is differentiable with respect to $\psi_0$, we can write
\beq \label{eq:Bayes}
	\rho_{\varphi \mid \psi} = \rho_{\varphi \mid i,\psi}^{1/2} \rho_{\varphi_0 \mid \psi_0} \rho_{\varphi \mid i,\psi}^{1/2}.
\eeq
We refer to the operator $\rho_{\varphi \mid i,\psi}$ as the \emph{conditional} spatial derivative of $\varphi$ with respect to $i$ and $\psi$. It is proportional to the spatial derivative $\rho_{i_{\varphi}^{\dagger} \mid i_{\psi}^{\dagger}}$ and reduces to the standard (relative) conditional density when such an object is well defined. Using eqn. \eqref{eq:Bayes} and \eqref{eq:relent} we can decompose the relative entropy as
\begin{flalign} \label{eq:dataprocessing}
	S(\varphi \parallel \psi) &= S(\varphi_0 \parallel \psi_0) + \langle \Omega_{\varphi}, \sum_{n = 0}^{\infty} \frac{c_n}{n!} \text{ad}_{\log \rho_{\varphi \mid i,\psi}}(\log \rho_{\varphi_0 \mid \psi_0}) \Omega_{\varphi} \rangle_{\psi} \nonumber \\
	&\equiv S(\varphi_0 \parallel \psi_0) - \langle \Omega_{\varphi}, C_{\varphi \mid i,\psi} \Omega_{\varphi} \rangle_{\psi}.
\end{flalign}
The latter term, which quantifies the difference between the relative entropy of $\varphi$ and $\psi$ when regarded as states on $A$ and when regarded as a states on $B$, is called the relative conditional entropy of $\varphi$ and $\psi$. If $\tau$ is a trace on $A$ which restricts to a trace $\tau_0 \equiv \tau \circ i$ on $B$, then a simple application of \eqref{eq:dataprocessing} and \eqref{eq:vNEntWRTtrace} tells us that
\beq
	S_{\tau}(\varphi) = S_{\tau_0}(\varphi) + \langle \Omega_{\varphi}, C_{\varphi \mid i,\tau} \Omega_{\varphi} \rangle_{\tau}.
\eeq
The latter term here is called simply the conditional entropy of $\varphi$ with respect to the trace $\tau$. 

\section{From Algebras to Path Integrals} \label{app: AlgPI}

A $C^*$ algebra is a complete normed vector space $A$ together with a multiplication $\mu: A \otimes A \rightarrow A$ and an involution $\star: A \rightarrow A$ such that
\beq
	\norm{\mu(\star(a) \otimes a)} = \norm{a}^2. 
\eeq
Hereafter, we shall use the conventional notation $\mu(a_1 \otimes a_2) = a_1 a_2$ and $\star(a) = a^*$. Our $C^*$ algebras are always assumed to be unital e.g. containing an identity element which we shall denote by $\op{1}_A$ or just $\op{1}$ if no possibility for confusion arises. 

A weight on a $C^*$ algebra is a positive, linear map $\varphi: A \rightarrow \mathbb{C}$. That is 
\beq
	\varphi(a + \lambda b) = \varphi(a) + \lambda \varphi(b), \qquad \forall a,b \in A_+ \; \lambda \in \mathbb{R}_+, \qquad \varphi(a^* a)  
\eeq
The domain of a weight $\varphi$, denoted by $\mathfrak{n}_{\varphi}$, is the set of elements $a \in A$ such that $\varphi(a^*a) < \infty$. If a weight is also unital in the sense that $\varphi(\op{1}_A) = 1$ it is called a state. We will denote the space of weights by $P(A)$ and the space of states by $S(A)$.

In a more down to earth way, $C^*$ algebras naturally arise as subsets of bounded operators acting on a Hilbert space. The $C^*$ algebra $A$ acts on a Hilbert space $\mathscr{H}$ via a representation $\pi: A \rightarrow B(\mathscr{H})$, where each $\pi(a)$ is itself a linear map from $\mathscr{H}$ to $\mathscr{H}$. From this point of view we can always construct states taking expectation values of operators with respect to normalized vectors $\xi \in \mathscr{H}$
\beq	 \label{Vector states}
	\omega_{\xi}(\pi(a)) \equiv \langle \xi, \pi(a) \xi \rangle_{\mathscr{H}}. 
\eeq

The observation \eqref{Vector states} may the the source of some confusion as it appears to suggest that the state $\omega_{\xi}$ is a pure state. To this point, we must bear in mind that in the algebraic picture notions of subsystem are generically encoded in algebraic inclusions rather than Hilbert spaces. A state $\psi \in S(A)$ is called pure with respect to $A$ if it cannot be written as a convex combination of any other states. That is, if $\psi = p \psi_1 + (1-p) \psi_2$ for some $p \in (0,1)$ and $\psi_1,\psi_2 \in S(A)$ then $\psi_1 = \psi_2 = \psi$. If conversely $\psi$ can be written as a convex combination of some other states it is called mixed. 

As $\omega_{\xi}$ is implemented by a vector on the `global' Hilbert space $\mathscr{H}$, it will be a pure state for the algebra $B(\mathscr{H})$. It may not, however, be a pure state when restricted to the subalgebra $\pi(A) \subset B(\mathscr{H})$. This is perhaps best illustrated through the notion of purification by which any mixed state on the algebra $A$ viewed abstractly as map $\psi: A \rightarrow \mathbb{C}$ can be implemented by a vector $\xi \in \mathscr{H}$ if $\mathscr{H}$ is `sufficiently large'. The formal statement of this result is related to the GNS construction which we reviewed in Appendix \ref{app: QCon}. 

With the above discussion in mind, let us note that if $A$ admits a tracial weight, $\tau \in P(A)$ such that $\tau(ab) = \tau(ba)$, then we can associate states with density operators:
\beq
	\varphi(a) = \tau(\rho_{\varphi} a), \qquad \rho_{\varphi} \in A_+. 
\eeq 
The density operator $\rho_{\varphi}$ can be regarded as a notion of state which is fully restricted to the subsystem. If $\varphi$ possesses a purification in the Hilbert space $\mathscr{H}$, e.g. so that $\varphi = \omega_{\xi_{\varphi}}\rvert_{\pi(A)}$ with $\xi_{\varphi} \in \mathscr{H}$, then $\rho_{\varphi}$ can morally be read as the density operator obtained by partial tracing $\ket{\xi_{\varphi}} \bra{\xi_{\varphi}}$ with respect to the `complement' of $A$.\footnote{The notion of partial tracing is really only well defined when there is a tensor factorization of the underlying Hilbert space. However, it provides a useful heuristic.} We shall denote the set of density operators on $A$ by $D(A)$.
 
\subsection{Coherent State Path Integrals}

To connect the algebraic picture to the path integral picture we will consider a special class of representations which we call \emph{coherent representations}. The central ingredient in a coherent representation is a special kind of Hilbert space called a \emph{reproducing kernel Hilbert space} (RKHS). There are several different ways to introduce the notion of an RKHS \cite{berlinet2004reproducing}, but the one which is most useful for our purposes here is the following: A RKHS $\mathscr{H}_{X}$ is a subspace of $L^2(X,d\mu)$ where $(X,\mu)$ is a measure space, which is generated by an overcomplete basis of states $\{c_x\}_{x \in X}$ whose inner product $\langle c_x, c_y \rangle_{X} \equiv K(x,y)$ defines a positive, symmetric kernel on $X$ satisfying the reproducing property
\beq
	K(x,y) = \int_{X} d\mu(z) K(x,z) K(z,y). 
\eeq
The RKHS $\mathscr{H}_{X} \subset L^2(X,d\mu)$ consists of those elements $\psi \in L^2(X,d\mu)$ which are compatible with the kernel in the sense that
\beq
	\psi(x) = \int_{X} d\mu(y) K(x,y) \psi(y). 
\eeq

The kernel $K$ allows us to endow $X$ with the structure of a symplectic Kahler manifold \cite{Odzijewicz1992Coherent}. The symplectic potential is given by $\Theta \equiv \iota^*(d_2 \ln K)$, and the symplectic form by $\Omega \equiv \iota^*(i d_1 d_2 \ln K)$. Here, $d_i$ are exterior derivatives on each copy of $X$ in the Cartesian product space $X \times X$ and $\iota: X \hookrightarrow X \times X$ is the diagonal embedding $x \mapsto (x,x)$. From this point of view, $\mathscr{H}_X$ can be interpreted as a geometric quantization of the symplectic manifold $(X,\Omega)$, with $c_x \in \mathscr{H}_X$ defining generalized coherent states \cite{Perelomov:1971bd}. The identification of $c_x$ with coherent states is underscored by the following path integral preparation of their inner products:
\beq \label{Inner Product Path Integral}
	\langle c_x, c_y \rangle_{X} \equiv \int_{\{\gamma: [0,1] \rightarrow X \; | \; \gamma(0) = y, \gamma(1) = x\}} \mathscr{D} \gamma \; e^{i \int_{0}^{1} \gamma^* \Theta}. 
\eeq

Eqn. \eqref{Inner Product Path Integral} is a generalization of the usual phase space path integral. Indeed, for the standard phase space $X = \mathbb{R}^2$, the symplectic potential is given by $\Theta = p dq$ and eqn. \eqref{Inner Product Path Integral} reproduces the kinetic part of the path integral. Noticeably absent is the Hamiltonian. From the point of view we are cultivated, the Hamiltonian is most naturally regarded as arising from the insertion of an operator into the inner product. 

We can define an explicit quantization map $\mathcal{Q}: C^{\infty}(X) \rightarrow B(\mathscr{H}_X)$ which intertwines the Poisson algebra of $X$ with the operator algebra on $\mathscr{H}_X$ in the sense of Dirac
\beq \label{Dirac}
	[\mathcal{Q}(f_1), \mathcal{Q}(f_2)] = i \mathcal{Q}(\{f_1,f_2\}_X). 
\eeq
While eqn. \eqref{Dirac} determines the commutation relation between operators, there are still the usual ordering ambiguities pertaining to the product. These can be dealt with within the path integral formalism by taking
\beq \label{Product form path}
	\mathcal{Q}(f) \mathcal{Q}(g) \equiv \mathcal{Q}(f \star g), \qquad \bigg(f \star g\bigg)(x) \equiv \int_{\{\gamma: (-\infty,\infty) \rightarrow X \; | \; \lim_{t \rightarrow \pm \infty} \gamma(t) = x\}} \mathscr{D} \gamma \; f \circ \gamma(1) g \circ \gamma(0) e^{i \int_{-\infty}^{\infty} \gamma^*\Theta}. 
\eeq
The latter equation in \eqref{Product form path} is a very formal notion of Kontsevich's deformation quantization \cite{Kontsevich:1997vb}.\footnote{Indeed, the $\star$ product is only well defined as an expansion in a formal parameter $\hbar$ which we have here set equal to one. This is only valid in the case of a strict deformation quantization, which should be attainable at least for  a subset of operators in our formulation.} Each $\mathcal{Q}(f)$ is a positive self-adjoint operator on $\mathscr{H}_X$ which can be exponentiated to a unitary operator $e^{i \mathcal{Q}(f)}$. We may then consider the matrix elements $\langle c_x, e^{i \mathcal{Q}(f)} c_y \rangle_X$, which are the generalization of the usual quantum mechanical propagator. In \cite{Odzijewicz1992Coherent} it is shown that, for quantizable functions, these matrix elements can too be given a path integral preparation
\beq \label{Matrix Elements}
	\langle c_x, e^{i \mathcal{Q}(f)t} c_y \rangle_X = \int_{\{\gamma: [0,t] \rightarrow X \; | \; \gamma(0) = y, \gamma(t) = x\}} \mathscr{D} \gamma \; e^{i \int_{0}^{t} (\gamma^* \Theta - \gamma^*f dt'}). 
\eeq	 
This is the phase space path integral with Hamiltonian function $f$. We will often use the notation
\beq
	\langle c_x, e^{i \mathcal{Q}(f) t} c_y \rangle_{X} = \int_{\{\gamma: [0,t] \rightarrow X \; | \; \gamma(0) = y, \gamma(t) = x\}} \mathscr{D} \gamma \; e^{i S^f[\gamma]}, \qquad S^f[\gamma] \equiv \int_{0}^{1} (\gamma^*\Theta - \gamma^*f dt).
\eeq
This makes the connection to the usual path integral interpretation clear.

If $\mathcal{Q}(f)$ has a gapped spectrum, we can also use it to define a state in the algebraic sense of an expectation value. First, we use the observation that 
\beq
	\xi_{f} \equiv \lim_{\tau \rightarrow \infty} e^{-\mathcal{Q}(f) \tau}(c_x) 
\eeq
defines a unique vector in $\mathscr{H}_{X}$ independently of the chosen coherent vector $c_x$. This is because $\lim_{\tau \rightarrow \infty} e^{-\mathcal{Q}(f)\tau}$ serves as a projection operator onto the `ground state' of the `Hamiltonian' $\mathcal{Q}(f)$. Then, for any other quantizable function $g \in C^{\infty}(X)$ we can define the state $\omega_{\xi_f}$ by the path integral
\begin{flalign} \label{State preparation}
	\omega_{\xi_f}(\mathcal{Q}(g)) &= \langle \lim_{\tau \rightarrow \infty} e^{-\mathcal{Q}(f) \tau} c_x, \mathcal{Q}(g) \lim_{\tau \rightarrow \infty} e^{-\mathcal{Q}(f) \tau} c_y \rangle_X \nonumber \\
	&= \int_{\{\gamma: (-\infty,\infty) \rightarrow X\}} \mathscr{D} \gamma \; g \circ \gamma(0) e^{i\int_{-\infty}^{\infty} (\gamma^*\Theta - i \gamma^* f dt)}. 
\end{flalign}
This can be interpreted as a `Euclidean signature' path integral, although we see that the complexification merely arises from the fact that the projector that defines the state is of the form $e^{i\mathcal{Q}(f)(i\tau)}$ rather that $e^{i\mathcal{Q}(f) t}$. We will use the notation
\beq
	\psi_f(\mathcal{Q}(g)) = \int_{\{\gamma: (-\infty,\infty) \rightarrow X\}} \mathscr{D} \gamma \; g \circ \gamma(0) e^{-I^{f}[\gamma]}
\eeq 
to again make contact with the standard formulae. We use the letter $I$ to distinguish from the action appearing in \eqref{Matrix Elements} and to remind us that state preparation in the path integral comes from a complexified contour. Again, the state $\omega_{\xi_f}$ is pure on the global algebra $B(\mathscr{H})$, but may be mixed when restricted to subalgebras therein. 

\subsection{Coherent Representation for $C^*$ Algebras}

A coherent representation of the $C^*$ algebra $A$ is a representation $\pi_X: A \rightarrow B(\mathscr{H}_X)$, where $\mathscr{H}_X \subset L^2(X,d\mu)$ is a RKHS. Let us denote by $A_X \equiv \pi_X(A) \cap \mathcal{Q}(C^{\infty}(X))$ the set of quantized operators from the phase space $X$ which also belong to the algebra $A$. Likewise, let us denote by $C_A^{\infty}(X) \equiv \mathcal{Q}^{-1}(A_X)$ the set of functions on $X$ which quantize to operators in $A$. 

For these operators we can utilize the above discussion to translate the computation of matrix elements and expectation values into familiar path integral expressions. Given an operator $a \in A_X$ we write $f_a \in C^{\infty}(X)$ for the phase space function such that $\pi_X(a) = e^{i \mathcal{Q}(f_a)}$. The matrix elements of $a$ in the coherent basis are therefore given by
\beq \label{Matrix elements algebra}
	\langle c_x, \pi_X(a) c_y \rangle_X = \int_{\{\gamma: [0,1] \rightarrow X \; | \; \gamma(0) = y, \gamma(1) = x\}} \mathscr{D} \gamma \; e^{i S^a[\gamma]}, \qquad S^a[\gamma] = S^{f_a}[\gamma]. 
\eeq 
Likewise, if $\psi \in S(A)$ is a state with vector representative $\xi_{\psi} = \xi_{f_{\psi}} \in \mathscr{H}_X$ then we can write
\beq \label{State prep algebra}
	\psi(\ln a) = \int_{\{\gamma: (-\infty,\infty) \rightarrow X\}} \mathscr{D} \gamma \; f_a \circ \gamma(0) e^{-I^{\psi}[\gamma]}, \qquad I^{\psi}[\gamma] = I^{f_{\psi}}[\gamma].
\eeq

It is enticing to regard $A$ as a quantization of the classical theory encoded in the phase space $X$. In many instances this interpretation is valid. However, more generally we would like to put forward the following interpretation: The algebra $A$ is a fundamentally quantum object. Each coherent representation $\pi_X: A \rightarrow B(\mathscr{H}_X)$ is simply a mathematical device for representing the standard operations within $A$ as functional integrals. We can think of these representations as `classicalizations' of the quantum theory encoded in $A$. Any given algebra $A$ may have many `classicalizations' which describe different coherent limits. The algebra $A$ may also possess some operators and states which do not have any classical correspondent. Exploring the interrelation between different coherent representations, and the question of which algebraic operators/states possess classical limits seems to be a novel approach to understanding the transit from quantum to classical physics. We plan to explore this in detail in future work. 


\subsection{Stinespring meets Feynman-Vernon} \label{sec: FV}

A significant ingredient in the investigations of the current paper is the notion of a quantum channel. A quantum channel $\alpha: A \rightarrow A$ defines a notion of open quantum system dynamics. To see this most clearly, we invoke the Stinespring dilation theorem. Suppose that $\pi: A \rightarrow B(\mathscr{H})$ is a representation of $A$. Stinespring's theorem\footnote{For a more rigorous statement of Stinespring's theorem see Appendix \ref{app: QCon}.} tells us that there exists an environment, modeled by the Hilbert space $\mathscr{H}_E$, and a representation $\Pi: A \rightarrow B(\mathscr{H} \otimes \mathscr{H}_E)$, such that
\beq
	\pi \circ \alpha(a) = V^{\dagger} \Pi(a) V
\eeq
with $V: \mathscr{H} \rightarrow \mathscr{H} \otimes \mathscr{H}_E$ an isometry. 

The Schrodinger dual of this statement is perhaps a more recognizable result. Supposing that $A$ admits a trace $\tau$ the pullback $\alpha^*: S(A) \rightarrow S(A)$ can be regarded as a map $\alpha^{\dagger}: D(A) \rightarrow D(A)$. Explicitly, 
\beq
	\alpha^*\psi(a) = \psi \circ \alpha(a) = \tau(\rho_{\psi} \alpha(a)) \equiv \tau(\alpha^{\dagger}(\rho_{\psi}) a) \iff \alpha^{\dagger}(\rho_{\psi}) = \rho_{\alpha^* \psi}.  
\eeq
That is, $\alpha^{\dagger}$ is the formal adjoint of the map $\alpha$ when $A$ is viewed as a Hilbert space with inner product $\langle a_1, a_2 \rangle_{\tau} \equiv \tau(a_1^* a_2)$.\footnote{Of course, this is the GNS Hilbert space of $A$ with respect to $\tau$, see again Appendix \ref{app: QCon}.} Then, Stinespring's theorem implies that the map $\alpha^{\dagger}$, which is completely positive and trace preserving, can be written as
\beq \label{Schrodinger Stinespring}
	\pi \circ \alpha^{\dagger}(\rho) = \text{ptr}_{\mathscr{H}_E}\bigg(U(\rho \otimes \mathbb{1}_{\mathscr{H}_E}) U^{\dagger}\bigg),
\eeq	 
where $U$ is a unitary operator on $\mathscr{H} \otimes \mathscr{H}_E$ and $\text{ptr}_{\mathscr{H}_E}$ is the partial trace. This is the standard statement that open quantum system evolution, in the Schrodinger picture, can be modeled as a unitary evolution on an enlarged system followed by a tracing out of the environmental degrees of freedom. 

Now, let us consider the case in which our algebra $A$ is coherently represented on an RKHS $\mathscr{H}_X$ and undergoes an open dynamics generated by the channel $\alpha$. Let us moreover assume that $\alpha$ can be spatially implemented on the Hilbert space $\mathscr{H}_X \otimes \mathscr{H}_E$, in which $\mathscr{H}_E$ is also an RKHS with underlying measure space $(E,d\mu_E)$. Then, using \eqref{Schrodinger Stinespring} we can give a path integral interpretation to the expectation value $\psi \circ \alpha$, where $\psi$ is the state on $A$ prepared by the path integral \eqref{State prep algebra}. 

The open dynamics are generated by first implementing the unitary $U = e^{i \mathcal{Q}_{X \times E}(F_{\alpha})}$ where here $F_{\alpha} \in C^{\infty}(X \times E)$, and then partial tracing over the environmental degrees of freedom. Here, we have used the notation $F_{\alpha}$ to remind ourselves that the form of $F_{\alpha}$ is determined by the channel $\alpha$. We have also amended the notation $\mathcal{Q}_{X}: C^{\infty}(X) \rightarrow B(\mathscr{H}_X)$ to signify the Poisson algebra it is quantizing. With these notations in place we can write
\beq \label{Feynman Vernon Path Integral}
	\psi \circ \alpha\bigg( \ln a \bigg) = \int_{\{\gamma = (\gamma_X,\gamma_E): (-\infty,\infty) \rightarrow X \times E\}} \mathscr{D} \gamma_X \mathscr{D} \gamma_E \; f_a \circ \gamma_X(0) e^{-I^{\psi,\alpha}[\gamma_X,\gamma_E]}. 
\eeq
The action appearing in \eqref{Feynman Vernon Path Integral} is given precisely by
\beq
	I^{\psi,\alpha}[\gamma_X,\gamma_E] = \int_{-\infty}^{\infty} (\gamma^* \Theta_{X \times E} - i\gamma_X^* f_{\psi} dt - \gamma^*F_{\alpha} dt).
\eeq
Where we have invoked \eqref{Matrix Elements} and \eqref{State preparation}. 

It is conceptually useful to split this action into the sum of three terms which (i) depend only on $\gamma_X$, (ii) depend only on $\gamma_E$, or (iii) encode an interaction between the two. This leads to the following path integral which prepares the state under the action of the channel:
\beq
	\psi \circ \alpha\bigg( \ln a \bigg) = \int_{\{\gamma = (\gamma_X,\gamma_E): (-\infty,\infty) \rightarrow X \times E\}} \mathscr{D} \gamma_E \mathscr{D} \gamma_X \; f_a \circ \gamma_X(0) e^{-\bigg(I^{f,\alpha}_X[\gamma_X] + I^{\alpha}[\gamma_E] + I^{\alpha}_{int}[\gamma_X,\gamma_E]\bigg)}.
\eeq	
This has the form of the standard Feynman-Vernon path integral for an open quantum system \cite{FEYNMAN1963118}. Indeed, the role of the quantum channel, in the path integral language, is to prepare the full action $S[\gamma_X,\gamma_E]$ for the combined system plus environment. 

By an analogous argument, we can also construct a path integral preparation of a quantum channel $\alpha: A \rightarrow B$ provided $B \subset A$ and assuming that $A$ is coherently represented on an RKHS $\mathscr{H}_X$ and $B$ is coherently represented on an RKHS $\mathscr{H}_Y \subset \mathscr{H}_X$ (equivalently $Y \subset X$). Again, we introduce an environment $\mathscr{H}_E$ such that
\beq
	\pi_B \circ \alpha^{\dagger}(\rho_B) = \text{ptr}_{\mathscr{H}_E} \bigg(U(\rho_B \otimes \mathbb{1}_{\mathscr{H}_E})U^{\dagger}\bigg),
\eeq
with $U$ a unitary on $\mathscr{H}_X \otimes \mathscr{H}_E$. Given a state $\psi \in S(B)$ we can then write
\begin{flalign} \label{FV Path 1}
	\psi \circ \alpha(\ln a) &= \int_{\{\gamma: (-\infty,\infty) \rightarrow X \times E\}} \mathscr{D} \gamma_E \mathscr{D} \gamma_X \; f_a \circ \gamma_X(0) e^{-I^{\psi,\alpha}[\gamma_X,\gamma_E]} \nonumber \\
	&= \int_{\{\gamma_X: (-\infty,\infty) \rightarrow X\}} \mathscr{D} \gamma_X \; f_a \circ \gamma_X(0) e^{-I^{\alpha^*\psi}_{X}[\gamma_X]}.
\end{flalign}
Here $I^{\alpha^*\psi}_X[\gamma_X]$ is the action on $X$ alone which is obtained after integrating out the environmental degrees of freedom. In this regard, we can view the map from $\psi \in S(B)$ to $\alpha^*\psi \in S(A)$ as changing the action from $I^{\psi}_Y[\gamma_Y]$ to $I^{\alpha^*\psi}_X[\gamma_X]$. Taking $X = Y \times Z$ we can generically write
\beq \label{FV Path 2}
	I^{\alpha^*\psi}_X[\gamma_Y,\gamma_Z] = I^{\psi}_Y[\gamma_Y] + I^{\alpha}_{int}[\gamma_Y,\gamma_Z]. 
\eeq
The latter quantity encodes the quantum channel as an interaction between the original system $Y$ and the new degrees of freedom $Z$ which complete $Y$ to the overall system $X$. 

\begin{table}[H]
\renewcommand{\arraystretch}{1.6}
\setlength{\tabcolsep}{8pt}
\begin{tabular}{|>{\raggedright\arraybackslash}p{0.22\textwidth}|p{0.72\textwidth}|}
\hline
\textbf{Product} &
\[
\pi_X(a_1)\pi_X(a_2) = \pi_X(a_{12})
\]
\[
f_{a_{12}}(x) =
\int_{\substack{\gamma: (-\infty,\infty)\to X \\ \lim_{t\to \pm \infty}\gamma(t)=x}}
\mathscr{D}\gamma \;
f_{a_1}(\gamma(1))\, f_{a_2}(\gamma(0)) \,
e^{i \int_{-\infty}^{\infty}\gamma^*\Theta}
\] \\
\hline

\textbf{Matrix Element} &
\[
\langle c_x, \pi_X(a) c_y \rangle_{X} =
\int_{\substack{\gamma: [0,1]\to X \\ \gamma(0)=y,\,\gamma(1)=x}}
\mathscr{D}\gamma \;
e^{i\int_0^1 \big(\gamma^*\Theta - \gamma^*f_a \, dt\big)}
\] \\
\hline

\textbf{State} &
\[
\psi(\ln a) =
\int_{\gamma: (-\infty,\infty)\to X}
\mathscr{D}\gamma \; f_a(\gamma(0)) \,
e^{-\int_{-\infty}^{\infty}\big(\gamma^*\Theta - i \gamma^* f_{\psi}\, d\tau\big)}
\] \\
\hline

\textbf{Channel} &
\[
\psi \circ \alpha(\ln a) =
\int_{\substack{\gamma: (-\infty,\infty) \\ \;\; \to X \times E}}
\mathscr{D}\gamma_E \, \mathscr{D}\gamma_X \; f_a(\gamma_X(0)) \,
e^{-\int_{-\infty}^{\infty}\big(
\gamma^* \Theta_{X \times E}
- i \gamma_X^* f_{\psi}\, d\tau - \gamma^* F_{\alpha}\, d\tau\big)}
\]
\\
\hline
\end{tabular}
\caption{Path integral preparations of products, matrix elements, states, and quantum channels for arbitrary algebras with coherent representations.}
\label{table: Algebra to Path Integral}
\end{table}

\pagebreak

\providecommand{\href}[2]{#2}\begingroup\raggedright\endgroup

\end{document}